\documentclass[reprint,aps,prb,twocolumn,amsmath,amssymb,showpacs,preprintnumbers]{revtex4-2}

\usepackage{graphicx}[]% Include figure files
\usepackage{dcolumn}% Align table columns on decimal point
\usepackage{bm}% bold math
\usepackage{multirow}

\usepackage{braket}
\usepackage{color}
\usepackage{ulem}
\usepackage{url}
\usepackage{hyperref}

\begin{document}

\title{
Uniform and Staggered electric axial moment in zigzag chain
}
\author{Satoru Hayami}
\affiliation{
Graduate School of Science, Hokkaido University, Sapporo 060-0810, Japan
}

\begin{abstract}
We theoretically investigate electronic orderings with the electric axial moment without breakings of both spatial inversion and time-reversal symmetries in the zigzag-chain system. 
Especially, we elucidate the role of the local odd-parity hybridization arising from locally noncentrosymmetric lattice structures based on symmetry and microscopic model analyses. 
We show that the odd-parity crystalline electric field gives rise to an effective cross-product coupling between the electric dipole and electric toroidal dipole, the latter of which corresponds to the electric axial moment. 
As a result, the staggered component of the electric axial moment is induced by applying an external electric field, while its uniform component is induced via the appearance of staggered electric dipole ordering. 
We also show that uniform electric quadrupole ordering accompanies uniform electric axial moment. 
Furthermore, we discuss transverse magnetization as a consequence of the orderings with the uniform electric axial moment. 
Our results extend the scope of materials exhibiting electric axial ordering to those with locally noncentrosymmetric lattice structures. 
\end{abstract}

\maketitle

\section{Introduction}

Polarity and axiality are important factors to determine physical properties in condensed matter physics. 
For example, a polar vector invariant under time-reversal operation is a source of electric polarization. 
When a system accommodates a polar vector, various parity-violating physical phenomena occur~\cite{Fu_PhysRevLett.115.026401, Kozii_PhysRevLett.115.207002, Venderbos_PhysRevB.94.180504}, such as antisymmetric spin-split band structure~\cite{rashba1960properties, Dresselhaus_Dresselhaus_Jorio}, spin Hall effect~\cite{murakami2003dissipationless, Murakami_PhysRevLett.93.156804, Sinova_PhysRevLett.92.126603, Fujimoto_doi:10.1143/JPSJ.75.083704,fujimoto2007fermi}, Edelstein effect~\cite{edelstein1990spin, Yip_PhysRevB.65.144508,Fujimoto_PhysRevB.72.024515,yoda2018orbital, Massarelli_PhysRevB.100.075136}, and nonlinear Hall effect~\cite{Sodemann_PhysRevLett.115.216806, Nandy_PhysRevB.100.195117}. 
When the time-reversal symmetry is fruther broken in the system with a polar vector, a magnetic toroidal dipole occurs~\cite{dubovik1975multipole,dubovik1990toroid,gorbatsevich1994toroidal, Spaldin_0953-8984-20-43-434203,van2007observation, cheong2018broken}, which exhibits a linear magnetoelectric effect~\cite{popov1999magnetic,schmid2001ferrotoroidics, EdererPhysRevB.76.214404, KhomskiiPhysics.2.20, zimmermann2014ferroic} and nonreciprocal transport~\cite{Sawada_PhysRevLett.95.237402,Miyahara_JPSJ.81.023712,Miyahara_PhysRevB.89.195145, tokura2018nonreciprocal}.  
Meanwhile, an axial vector that inverts under time-reversal operation gives rise to magnetization, which becomes the origin of the anomalous Hall effect in collinear~\cite{Ye_PhysRevLett.83.3737, Solovyev_PhysRevB.55.8060, Chen_PhysRevB.106.024421, Naka_PhysRevB.102.075112, Hayami_PhysRevB.103.L180407}, noncollinear~\cite{Tomizawa_PhysRevB.80.100401,Chen_PhysRevLett.112.017205,nakatsuji2015large,Suzuki_PhysRevB.95.094406,Chen_PhysRevB.101.104418}, and noncoplanar magnets~\cite{Ohgushi_PhysRevB.62.R6065, Shindou_PhysRevLett.87.116801, Nagaosa_RevModPhys.82.1539}. 

Recently, an electric axiality, which has the opposite time-reversal parity to the magnetization, has attracted growing interest~\cite{Hlinka_PhysRevLett.113.165502, Hlinka_PhysRevLett.116.177602}, since the direct observation of its electronic ordering termed as ferro-axial (or ferro-rotational) ordering in RbFe(MoO$_4$)$_2$~\cite{jin2020observation, Hayashida_PhysRevMaterials.5.124409} and NiTiO$_3$~\cite{hayashida2020visualization, Hayashida_PhysRevMaterials.5.124409, yokota2022three}. 
Owing to the different time-reversal parity, a ferro-axial ordered state exhibits qualitatively different physical phenomena from the conventional ferromagnetic ordering~\cite{cheong2021permutable, Hayami_PhysRevB.106.144402, cheong2022linking, hayami2023planar}, such as antisymmetric thermopolarization~\cite{Nasu_PhysRevB.105.245125}, longitudinal spin current generation~\cite{Roy_PhysRevMaterials.6.045004, Hayami_doi:10.7566/JPSJ.91.113702}, and nonlinear transverse magnetization~\cite{inda2023nonlinear}. 
However, materials to be identified as ferro-axial ordering in experiments are much smaller than those as ferromagnetic ordering: Co$_3$Nb$_2$O$_8$~\cite{Johnson_PhysRevLett.107.137205}, CaMn$_7$O$_{12}$~\cite{Johnson_PhysRevLett.108.067201}, Ca$_5$Ir$_3$O$_{12}$~\cite{Hasegawa_doi:10.7566/JPSJ.89.054602, hanate2021first, hayami2023cluster, hanate2023space}, BaCoSiO$_4$~\cite{Xu_PhysRevB.105.184407}, K$_2$Zr(PO$_4$)$_2$~\cite{yamagishi2023ferroaxial}, Na$_2$Hf(BO$_3$)$_2$~\cite{nagai2023chemicalSwitching}, and Na-superionic conductors~\cite{nagai2023chemical}. 
In order to extend the scope of candidate materials, it is important to propose various potential situations to accommodate the electric axial moment including not only the ferro-axial moment but also the antiferro-axial moment. 
Especially, the analysis of the electric axial moment in a simple lattice structure is desired in order to understand its fundamental nature. 

In the present study, we investigate a fundamental situation, where the uniform and staggered electric axial moments can emerge, by focusing on a zigzag-chain structure with local inversion symmetry breaking. 
The aim of this study is to clarify the relationship between the electric axial moment and the locally odd-parity crystalline electric field that arises from the lack of an inversion center at each lattice site in order to further understand the behavior of the electric axial moment, which will be applied to not only a vast of zigzag-chain materials but also materials with similar lattice structures, such as honeycomb and diamond structures. 
To extract an essence, we analyze a minimal four-orbital model which incorporates the effect of the $s$--$p$ hybridization and atomic spin--orbit coupling in the one-dimensional zigzag chain. 
We find that the site-dependent spin--orbit interaction originating from the local parity mixing in the zigzag chain gives rise to an effective cross-product coupling between an electric field and staggered electric axial moment. 
Moreover, in analogy to staggered antiferromagnetic ordering with uniform magnetic toroidal dipole~\cite{Yanase_JPSJ.83.014703, Hayami_doi:10.7566/JPSJ.84.064717}, we show that staggered electric dipole ordering in the zigzag chain accompanies uniform electric toroidal dipole corresponding to the uniform component of the electric axial moment. 
We also show that uniform electric quadrupole ordering is another one to accompany the uniform electric axial moment. 
In both cases, we demonstrate the emergence of the transverse magnetization characteristic of the uniform electric axial state, which can be observed in experiments.

The rest of this paper is organized as follows. 
In Sec.~\ref{sec: Ferroaxial moment under locally asymmetric crystal field}, we present a relationship between the electric axial moment and odd-parity crystalline electric field based on the symmetry and augmented multipole analyses. 
After introducing a minimal tight-binding model in Sec.~\ref{sec: Model}, we show two situations to activate the electric axial moment in the zigzag chain: One is the application of the external electric field to cause the staggered electric axial moment in Sec.~\ref{sec: Staggered ferroaxial moment induced by external electric field} and the other is staggered electronic orderings to induce the uniform one in Sec.~\ref{sec: Uniform ferroaxial moment induced by electronic ordering}. 
Section~\ref{sec: Summary} is devoted to a summary of this paper.

\section{Electric axial moment under locally asymmetric crystal field}
\label{sec: Ferroaxial moment under locally asymmetric crystal field}

\begin{figure}[tb!]
\begin{center}
\includegraphics[width=1.0\hsize]{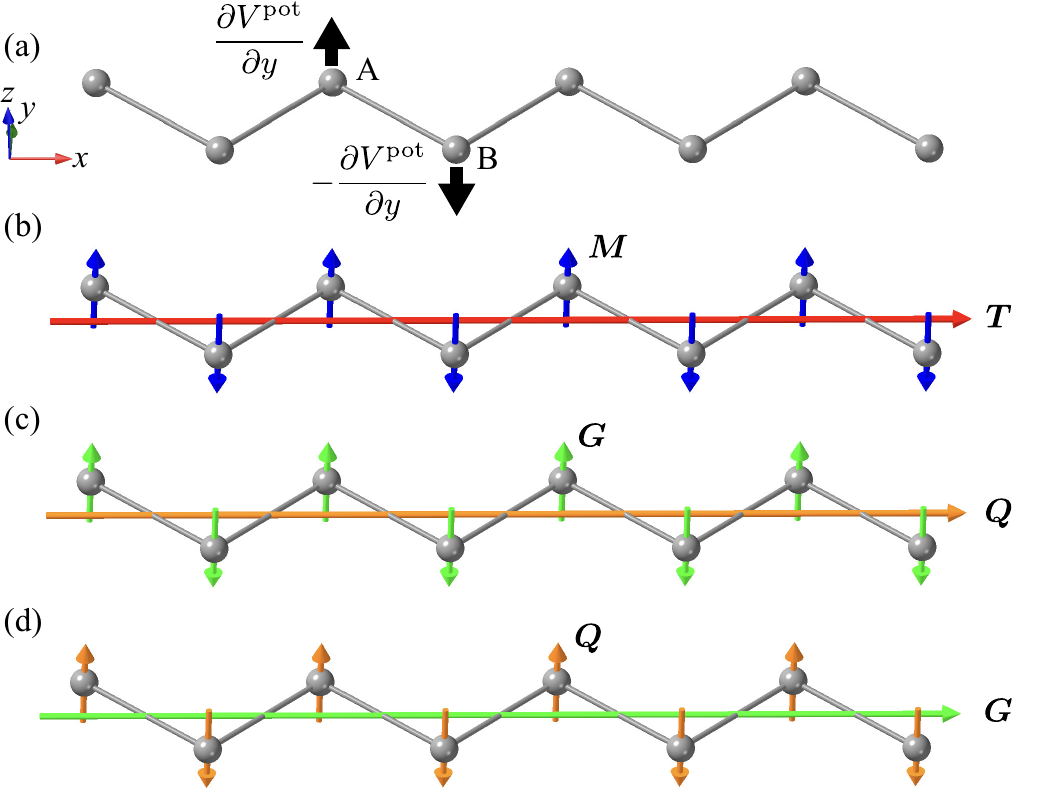} 
\caption{
\label{fig: ponti} 
(a) Zigzag chain consisting of two sublattices, A and B. 
The black arrows represent the direction of sublattice-dependent potential gradient $\pm \partial V^{\rm pot} /{\partial y}$ corresponding to the odd-parity crystalline electric field.  
(b) Staggered alignment of magnetic dipole $\bm{M}$ accompanying the magnetic toroidal dipole $\bm{T}$ in the zigzag chain.  
(c) Staggered alignment of the electric toroidal dipole $\bm{G}$ accompanying the electric dipole $\bm{Q}$. 
(d) Staggered alignment of $\bm{Q}$ with uniform $\bm{G}$. 
}
\end{center}
\end{figure}

We discuss the relationship between the electric axial moment and local odd-parity crystalline electric field by introducing four types of multipoles (electric, magnetic, electric toroidal, and magnetic toroidal multipoles) with different spatial-inversion and time-reversal parities~\cite{kusunose2022generalization}; the electric (magnetic toroidal) multipole corresponds to the time-reversal even (odd) polar tensor, while the electric toroidal (magnetic) multipole corresponds to the time-reversal even (odd) axial tensor. 
We consider the one-dimensional zigzag chain along the $x$ direction in Fig.~\ref{fig: ponti}(a), which belongs to the point group $D_{\rm 2h}$. 
Although there is inversion symmetry around the bond center between A and B sublattices, the local inversion symmetry is broken at each sublattice; the site symmetry belongs to the polar point group $C_{\rm 2v}$. 
Accordingly, the linear potential gradient along the $y$ direction occurs in an opposite direction in each sublattice: $\partial V^{\rm pot} /{\partial y}$ for the A sublattice, while $- \partial V^{\rm pot} /{\partial y}$ for the B sublattice. 
Such a potential gradient leads to the sublattice-dependent hybridization between the orbitals with different spatial parity, and results in a sublattice-dependent antisymmetric spin--orbit interaction by considering the combination with the atomic spin--orbit coupling~\cite{Yanase_JPSJ.83.014703}. 

The sublattice-dependent antisymmetric spin--orbit interaction has been recently studied in various contexts in condensed matter physics, since it becomes the origin of various electronic orderings, such as odd-parity magnetic multipole orderings~\cite{Yanase_JPSJ.83.014703, Hayami_doi:10.7566/JPSJ.84.064717, Hayami_PhysRevB.90.081115,hayami2016emergent, Watanabe_PhysRevB.96.064432, Hayami_PhysRevB.104.045117, Kirikoshi_PhysRevB.107.155109}, skyrmion crystals~\cite{Hayami_PhysRevB.105.014408, Hayami_PhysRevB.105.184426, hayami2022square, lin2021skyrmion}, and exotic superconducting states~\cite{Maruyama_doi:10.1143/JPSJ.81.034702, Goryo_PhysRevB.86.100507, Yoshida_PhysRevB.86.134514, yoshida2013parity, Yoshida_doi:10.7566/JPSJ.82.074714, sigrist2014superconductors, Yoshida_PhysRevLett.115.027001, Sumita_PhysRevB.93.224507, Ishizuka_PhysRevB.98.224510, Nogaki_PhysRevResearch.3.L032071, Mockli_PhysRevB.104.134517, fischer2023superconductivity}. 
For example, the relevance with magnetic odd-parity multipole orderings is understood from the symmetry correspondence as follows: 
\begin{align}
\label{eq: Vpot_MT}
\frac{\partial V^{\rm pot}}{\partial y} \leftrightarrow (\bm{M} \times \bm{T})^y, 
\end{align}
where $\bm{M}$ and $\bm{T}$ represent the magnetic dipole and magnetic toroidal dipole, respectively.  
The former corresponds to the time-reversal odd axial vector, while the latter corresponds to the time-reversal odd polar vector. 
When the staggered structure of the potential gradient $\partial V^{\rm pot}_{\rm stag}/\partial y = \partial V^{\rm pot}_{\rm A}/\partial y-\partial V^{\rm pot}_{\rm B}/\partial y$ (the subscript represents the sublattice index) is considered with the zigzag structure in mind, Eq.~(\ref{eq: Vpot_MT}) is rewritten as 
\begin{align}
\label{eq: Vpot_MT2}
\frac{\partial V^{\rm pot}_{\rm stag}}{\partial y} \leftrightarrow (\bm{M}_{\rm stag} \times \bm{T}_{\rm uni})^y, 
\end{align}
where $\bm{M}_{\rm stag}=\bm{M}_{\rm A} - \bm{M}_{\rm B}$ and $\bm{T}_{\rm uni}=\bm{T}_{\rm A} + \bm{T}_{\rm B}$. 
Equation~(\ref{eq: Vpot_MT2}) means that the staggered potential gradient induces an effective cross-product coupling between the staggered magnetic dipole and uniform magnetic toroidal dipole. 
Thus, the staggered antiferromagnetic ordered state with the $z$-spin polarization activates the uniform magnetic toroidal dipole along the $x$ direction in Fig.~\ref{fig: ponti}(b), which results in $\bm{T}$-related physical phenomena, such as magnetoelectric effect~\cite{cysne2021orbital}, nonlinear transport~\cite{Suzuki_PhysRevB.105.075201, Yatsushiro_PhysRevB.105.155157}, nonlinear spin Hall effect~\cite{Kondo_PhysRevResearch.4.013186, Hayami_PhysRevB.106.024405}, and nonreciprocal magnon excitations~\cite{Miyahara_JPSJ.81.023712, Miyahara_PhysRevB.89.195145, Hayami_doi:10.7566/JPSJ.85.053705, Takashima_PhysRevB.98.020401, Matsumoto_PhysRevB.101.224419, Matsumoto_PhysRevB.104.134420, Hayami_PhysRevB.105.014404}. 
Besides, since the magnetic toroidal dipole is the same symmetry as the electric current, the staggered component of the magnetization can be induced when applying an electric current in the paramagnetic state, which is similar to the Edelstein effect in noncentrosymmetric crystals~\cite{Yanase_JPSJ.83.014703, Hayami_doi:10.7566/JPSJ.84.064717}. 
These physical phenomena have been discussed in other similar locally noncentrosymmetric lattice structures, such as honeycomb~\cite{Kane_PhysRevLett.95.226801, Hayami_PhysRevB.90.081115, hayami2016emergent, yanagi2017optical, Yanagi_PhysRevB.97.020404}, diamond~\cite{Fu_PhysRevLett.98.106803, Hayami_PhysRevB.97.024414, Ishitobi_doi:10.7566/JPSJ.88.063708}, and bilayer~\cite{hitomi2014electric, hitomi2016electric, yatsushiro2020odd, Yatsushiro_PhysRevB.102.195147} structures, which provides useful information to explore and understand the metallic materials with odd-parity magnetic multipoles, such as UNi$_4$B~\cite{saito2018evidence, Yanagisawa_PhysRevLett.126.157201, ota2022zero} and Ce$_3$TiBi$_5$~\cite{motoyama2018magnetic, shinozaki2020magnetoelectric,shinozaki2020study, Hayami_doi:10.7566/JPSJ.91.123701}.  

In analogy to Eq.~(\ref{eq: Vpot_MT}), the odd-parity potential gradient is symmetrically related to electric-type multipoles with time-reversal even as 
\begin{align}
\frac{\partial V^{\rm pot}}{\partial y} \leftrightarrow (\bm{G} \times \bm{Q})^y, 
\end{align}
where $\bm{G}$ and $\bm{Q}$ stand for the electric toroidal dipole and electric dipole, respectively; $\bm{G}$ ($\bm{Q}$) corresponds to the electric axial (polar) vector with the opposite time-reversal parity to $\bm{M}$ ($\bm{T}$). 
In the zigzag-chain structure with the staggered potential gradient, the above expression is rewritten as 
\begin{align}
\label{eq: Vpot_ET_stag}
\frac{\partial V^{\rm pot}_{\rm stag}}{\partial y} \leftrightarrow (\bm{G}_{\rm stag} \times \bm{Q}_{\rm uni})^y, 
\end{align}
where $\bm{G}_{\rm stag}=\bm{G}_{\rm A} - \bm{G}_{\rm B}$ and $\bm{Q}_{\rm uni}=\bm{Q}_{\rm A} + \bm{Q}_{\rm B}$.
Thus, the staggered alignment of the $z$-directional electric axial moment, i.e., $G^z_{\rm stag} \neq 0$, induces the uniform electric dipole (electric polarization) along the $x$ direction, as shown in Fig.~\ref{fig: ponti}(c). 
Conversely, when the electric field along the $x$ direction is applied in the paramagnetic state, one can expect the $z$ component of the staggered electric axial moment, since the electric field is the same symmetry as $\bm{Q}_{\rm uni}$. 

Moreover, one can further rewrite the expression in Eq.~(\ref{eq: Vpot_ET_stag}) while preserving the symmetry as follows: 
\begin{align}
\label{eq: Vpot_ET_stag2}
\frac{\partial V^{\rm pot}_{\rm stag}}{\partial y} \leftrightarrow (\bm{G}_{\rm uni} \times \bm{Q}_{\rm stag})^y, 
\end{align}
where $\bm{G}_{\rm uni}=\bm{G}_{\rm A} + \bm{G}_{\rm B}$ and $\bm{Q}_{\rm stag}=\bm{Q}_{\rm A} - \bm{Q}_{\rm B}$.
This correspondence indicates that the staggered electric dipole ordering in the $z$ component accompanies the uniform axial moment along the $x$ direction, as shown in Fig.~\ref{fig: ponti}(d). 
In this way, the staggered (uniform) alignment of the electric dipole leads to the uniform (staggered) alignment of the electric axial moment in the zigzag-chain system.

\section{Model}
\label{sec: Model}

In order to demonstrate the above effective cross-product coupling between the electric dipole and electric axial moments beyond the symmetry argument, we investigate a minimal fundamental model in the zigzag chain. 
To incorporate the effect arising from the absence of the local inversion symmetry like  the odd-parity hybridization, the atomic spin--orbit coupling, and dipole degrees of freedom in terms of $\bm{Q}$ and $\bm{G}$, we consider the $s$-$p$ hybridized model consisting of four orbitals ($s$, $p_x$, $p_y$, $p_z$) in the zigzag chain, which is given by 
\begin{align}
\label{eq: Ham}
\mathcal{H}&= \mathcal{H}^t+\mathcal{H}^{\rm pot}+\mathcal{H}^{\rm SOC}+\mathcal{H}^{{\rm odd}},  \nonumber \\
\mathcal{H}^t  &=-\sum_{ij\alpha\alpha' \sigma}(t^{\alpha \alpha'}_{ij} c^{\dagger}_{i\alpha'\sigma} c_{j\alpha\sigma} + {\rm H.c.}),  \\
\mathcal{H}^{\rm pot}  &=\Delta \sum_{i \sigma} c^{\dagger}_{i s \sigma} c_{i s\sigma},  \\
\mathcal{H}^{{\rm SOC}}&= \frac{\lambda}{2} \sum_{i \tilde{\alpha} \tilde{\alpha}' \sigma \sigma'} c^{\dagger}_{i \tilde{\alpha} \sigma} H^{\rm SOC} c_{i \tilde{\alpha}' \sigma'}, \\ 
\label{eq: Hodd}
\mathcal{H}^{{\rm odd}}&= -V \sum_{i \sigma} p_i (c^{\dagger}_{i s \sigma} c_{i p_y \sigma}+ {\rm H.c.}), 
\end{align}
where $c^{\dagger}_{i\alpha\sigma}$ and $c_{i\alpha\sigma}$ stand for the creation and annihilation fermion operators at site $i$, orbital $\alpha=s$, $p_x$, $p_y$, and $p_z$, and spin $\sigma$, respectively. 
The total Hamiltonian consists of the hopping Hamiltonian $\mathcal{H}^t$, onsite-potential Hamiltonian $\mathcal{H}^{\rm pot}$, spin--orbit-coupling Hamiltonian $\mathcal{H}^{{\rm SOC}}$, and odd-parity mixing Hamiltonian $\mathcal{H}^{{\rm odd}}$. 
$\mathcal{H}^{t}$ represents the nearest-neighbor hopping between the A and B sublattices; we use four Slater-Koster parameters $t_{ss\sigma}$, $t_{pp\sigma}$, $t_{pp\pi}$, and $t_{sp\sigma}$ for $t^{\alpha \alpha'}_{ij}$ by setting the positions of the A and B sublattices as $(0,0,0)$ and $(a/2, -a/2, 0)$; $a$ is the lattice constant along the $x$ direction and we take $a=1$ as the length unit. 
The hoppings are parametrized as $(t_{ss\sigma}, t_{pp\sigma}, t_{pp\pi}, t_{sp\sigma})=(1, -0.8, 0.4, -0.6)$; we set $t_{ss\sigma}$ as the energy unit of the total Hamiltonian in Eq.~(\ref{eq: Ham}). 
$\mathcal{H}^{\rm pot}$ represents the atomic energy level for the $s$ orbital, which is lowered from the $p$-orbital level by $\Delta=-2$.
For simplicity, we ignore other even-parity crystalline electric fields, which split into the three $p$ orbitals. 
$\mathcal{H}^{{\rm SOC}}$ with $\tilde{\alpha}=p_x$, $p_y$, and $p_z$ represents the atomic spin--orbit coupling for three $p$ orbitals with the orbital angular momenta $l=1$, whose $6\times 6$ matrix is given by 
\begin{align}
 H^{\rm SOC} =  \left(
 \begin{array}{ccc}
 0 & -i \sigma^z & i \sigma^y \\
 i \sigma^z & 0 & -i \sigma^x \\
 -i \sigma^y & i \sigma^x & 0
 \end{array}
 \right), 
\end{align}
where $\sigma^{\mu}$ is the $\mu$ component of the Pauli matrix in spin space. 
We take the magnitude of the spin--orbit coupling as $\lambda=0.5$. 
$\mathcal{H}^{{\rm odd}}$ represents an odd-parity crystalline electric field depending on the sublattices, which arises from the absence of local inversion symmetry. 
Since the local potential gradient appears along the $y$ direction, as shown in Fig.~\ref{fig: ponti}(a), the local $s$--$p_y$ hybridization so as to mix different parities occurs. 
Owing to the presence of the global inversion symmetry around the nearest-neighbor bond center, the sign of the hybridization is opposite for the A and B sublattices; $p_i = + 1 (-1)$ for the A (B) sublattice. 
We set $V=0.3$. 

To discuss the cross-product coupling between $\bm{Q}$ and $\bm{G}$ at the microscopic level, we define their operator expressions. 
Based on the augmented multipole description~\cite{hayami2018microscopic, kusunose2020complete, Yatsushiro_PhysRevB.104.054412, Kusunose_PhysRevB.107.195118}, their expressions for $\bm{Q}_i=(Q^x_i, Q^y_i, Q^z_i)$ and $\bm{G}_i=(G^x_i, G^y_i, G^z_i)$ at site $i$ are given by 
\begin{align}
Q^x_i=&\sum_{\sigma} (c^{\dagger}_{i s \sigma} c_{i p_x \sigma}+ {\rm H.c.}), \\
Q^y_i=&\sum_{\sigma} ( c^{\dagger}_{i s \sigma} c_{i p_y \sigma}+ {\rm H.c.}), \\
\label{eq: Qz}
Q^z_i=& \sum_{\sigma} (c^{\dagger}_{i s \sigma} c_{i p_z \sigma}+ {\rm H.c.}), \\
\label{eq: Gx}
G^x_i=&  -i c^{\dagger}_{i p_z \uparrow} c_{i p_x \uparrow}  + c^{\dagger}_{i p_y \downarrow} c_{i p_x \uparrow} 
\nonumber \\ 
&-c^{\dagger}_{i p_x \downarrow} c_{i p_y \uparrow}  +i c^{\dagger}_{i p_z \downarrow} c_{i p_x \downarrow} + {\rm H.c.},\\ 
\label{eq: Gy}
G^y_i=&  i c^{\dagger}_{i p_y \downarrow} c_{i p_x \uparrow}  -i c^{\dagger}_{i p_z \uparrow} c_{i p_y \uparrow} 
\nonumber \\ 
&-i c^{\dagger}_{i p_x \downarrow} c_{i p_y \uparrow}  +i c^{\dagger}_{i p_z \downarrow} c_{i p_y \downarrow} + {\rm H.c.},\\ 
\label{eq: Gz}
G^z_i=&  i c^{\dagger}_{i p_z \downarrow} c_{i p_x \uparrow}  - c^{\dagger}_{i p_z \downarrow} c_{i p_y \uparrow} 
\nonumber \\ 
&-i c^{\dagger}_{i p_x \downarrow} c_{i p_z \uparrow}  + c^{\dagger}_{i p_y \downarrow} c_{i p_z \uparrow} + {\rm H.c.} 
\end{align}
From these expressions, one finds that $\bm{Q}_i$ corresponds to the real $s$-$p$ hybridization without spin dependence, while $\bm{G}_i$ corresponds to the outer product of the orbital angular momentum and spin operators ($\bm{l}_i \times \bm{\sigma}_i$)~\cite{Hayami_PhysRevB.98.165110}, each of which is represented by 
\begin{align}
l^x_i=&\sum_{\sigma} (i c^{\dagger}_{i p_z \sigma} c_{i p_y \sigma}+ {\rm H.c.}),  \\
l^y_i=&\sum_{\sigma} (i c^{\dagger}_{i p_x \sigma} c_{i p_z \sigma}+ {\rm H.c.}),  \\
l^z_i=&\sum_{\sigma} (i c^{\dagger}_{i p_y \sigma} c_{i p_x \sigma}+ {\rm H.c.}),  \\
\sigma^x_i=&\sum_{\sigma \sigma' \alpha} c^{\dagger}_{i \alpha \sigma} \sigma^x_{\sigma\sigma'} c_{i \alpha \sigma'}, \\
\sigma^y_i=&\sum_{\sigma \sigma' \alpha} c^{\dagger}_{i \alpha \sigma} \sigma^y_{\sigma\sigma'} c_{i \alpha \sigma'}, \\
\sigma^z_i=&\sum_{\sigma \sigma' \alpha} c^{\dagger}_{i \alpha \sigma} \sigma^z_{\sigma\sigma'} c_{i \alpha \sigma'}, 
\end{align}
where the spin operator $\bm{s}_i$ is defined by $\bm{s}_i=\bm{\sigma}_{i}/2$. 
It is noted that the staggered potential gradient in Eq.~(\ref{eq: Hodd}) is also described by the staggered alignment of the $y$ component of the electric dipole.

\section{Staggered axial moment in an external electric field}
\label{sec: Staggered ferroaxial moment induced by external electric field}

\begin{figure}[tb!]
\begin{center}
\includegraphics[width=1.0\hsize]{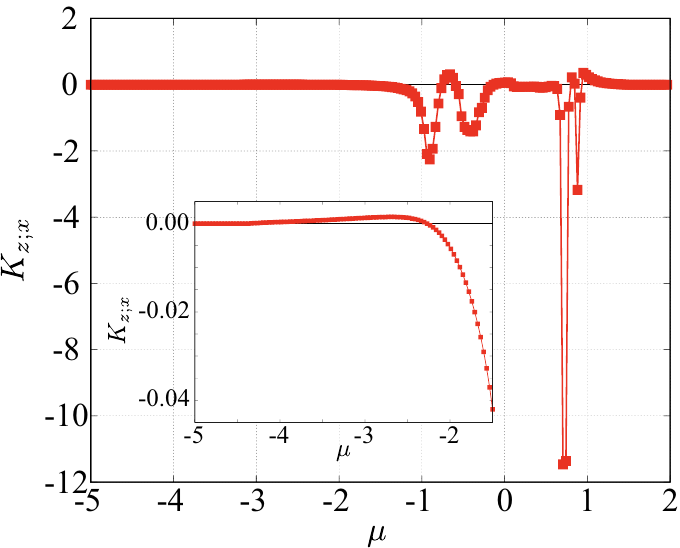} 
\caption{
\label{fig: staggered_res} 
Chemical potential $\mu$ dependence of the electric-field-induced staggered electric axial moment $K_{z;x}$. 
The inset shows the data in the low-filling region for $-5 \leq \mu \leq -1.5$. 
}
\end{center}
\end{figure}

\begin{figure}[tb!]
\begin{center}
\includegraphics[width=1.0\hsize]{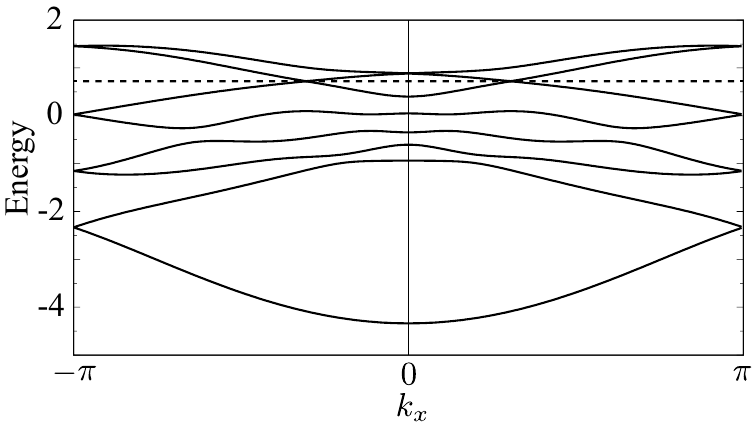} 
\caption{
\label{fig: band} 
Band dispersion of the model in Eq.~(\ref{eq: Ham}). 
The horizontal dashed line represents the energy, where $K_{z;x}$ in Fig.~\ref{fig: staggered_res} is maximized. 
}
\end{center}
\end{figure}

We examine the effective cross-product coupling between the staggered electric toroidal dipole $\bm{G}$ and the uniform electric dipole $\bm{Q}$ in Eq.~(\ref{eq: Vpot_ET_stag}). 
As discussed in Sec.~\ref{sec: Ferroaxial moment under locally asymmetric crystal field}, the $z$ component of the staggered electric axial moment is induced by applying the external electric field along the $x$ direction parallel to the zigzag-chain direction. 
To microscopically evaluate such a cross-product coupling, we calculate the correlation function between them based on the Kubo formula, which is given by 
\begin{align}
K_{z;x} =\frac{1}{i N} \sum_{m,n,\bm{k}} \frac{f(\varepsilon_{n \bm{k}})-f(\varepsilon_{m \bm{k}})}{\varepsilon_{n \bm{k}}-\varepsilon_{m \bm{k}}} 
\frac{
 G_{z, \bm{k}}^{nm} J_{x,\bm{k}}^{mn} }{
\varepsilon_{n \bm{k}}-\varepsilon_{m \bm{k}}+i \delta}, 
\label{Eq: Kzx}
\end{align}
where $N$ is the number of sites, $f(\varepsilon)$ is the Fermi distribution function, $\varepsilon_{n \bm{k}}$ and $| n \bm{k} \rangle$ are the $n$th eigenvalue and eigenstate of $\mathcal{H}$, respectively. 
$ G_{z,\bm{k}}^{nm}=\langle n \bm{k} | p_\eta G^{z}_\eta | m\bm{k} \rangle$ [$p_\eta = + 1 (-1)$ for the A (B) sublattice] and $J_{x,\bm{k}}^{mn}=\langle m \bm{k} |J_{x} | n\bm{k} \rangle$ are the matrix elements of staggered electric toroidal dipole operator and the current operator
$J_x = (e/\hbar)\partial \mathcal{H}
/\partial k_x$, respectively. 
We set $e/\hbar=1$, the temperature $T=0.01$, and the broadning factor $\delta=10^{-5}$. 
$K_{z;x}$ represents the linear-response function where the $z$ component of the staggered electric axial moment is induced by the external electric field along the $x$ direction. 

Figure~\ref{fig: staggered_res} shows the chemical potential $\mu$ dependence of $K_{z;x}$ for the model in Eq.~(\ref{eq: Ham}); the regions for $\mu \lesssim -4.6$ and $\mu \gtrsim 1.65$ correspond to those of zero electron filling and full filling, respectively. 
We also show the data in the low-filling region in the inset of Fig.~\ref{fig: staggered_res}. 
As found in the behavior of $K_{z;x}$, it becomes nonzero for all $\mu$. 
Thus, the zigzag-chain system under the locally noncentrosymmetric lattice structure exhibits instability toward the staggered electric axial moment when the electric field is applied, as expected from the symmetry argument in Sec.~\ref{sec: Ferroaxial moment under locally asymmetric crystal field}. 

In order to extract the essential model parameters to cause nonzero $K_{z;x}$ at the qualitative level, we perform the expansion method for the linear response function following the manner in Ref.~\cite{Oiwa_doi:10.7566/JPSJ.91.014701}. 
As a result, we find that the spin--orbit coupling $\lambda$ is included in the expansion of $K_{z;x}$ at any order, which means that $\lambda$ is essential for $K_{z;x}$. 
In addition, we find that all the orders in the expansion vanish when both the $s$-$p$ hybridization $V$ and the hopping between the $s$-$p$ orbitals $t_{sp\sigma}$ are zero, i.e., $K_{z;x}=0$ for $V=t_{sp \sigma}=0$; the odd-parity hybridization also plays an important role in inducing $K_{z;x}$. 
Such a behavior is also confirmed by directly performing a numerical evaluation of Eq.~(\ref{Eq: Kzx}). 

At the quantitative level, $K_{z;x}$ is enhanced at some particular $\mu$, as shown in Fig.~\ref{fig: staggered_res}; the largest response is obtained at $\mu \simeq 0.72$. 
Since the dominant process for $K_{z;x}$ is the interband one with $m \neq n$, one notices that the small energy difference in the denominator in Eq.~(\ref{Eq: Kzx}) tends to enhance $K_{z;x}$. 
Indeed, as shown in the band structure in Fig.~\ref{fig: band}, the chemical potential that maximizes $K_{z;x}$ lies in the small band gap denoted by the horizontal dashed line in Fig.~\ref{fig: band}, although it is difficult to see the band gap owing to the small energy difference. 
Thus, the small gap structure in the band dispersion is desired to obtain a large response of $K_{z;x}$. 

Let us discuss the difference from different staggered responses in the zigzag-chain system, where the $z$ component of the staggered magnetization is induced by an external electric current parallel to the chain direction in Fig.~\ref{fig: ponti}(b)~\cite{Yanase_JPSJ.83.014703, Hayami_doi:10.7566/JPSJ.84.064717}. 
Although such a response can be calculated by replacing $G^{z}_\eta$ with $\sigma^z_\eta$ in Eq.~(\ref{Eq: Kzx}), the dominant process becomes the intraband process related to $\delta$, which is different from the interband process in the present $K_{z;x}$. 
Thus, $K_{z;x}$ becomes nonzero for both metals and insulators, while it appears only for metals in the case of staggered magnetization.

\section{Uniform axial moment under electronic ordering}
\label{sec: Uniform ferroaxial moment induced by electronic ordering}

We discuss the situation where the uniform electric axial moment is induced by two types of electronic orderings: One is the staggered electric dipole ordering in Sec.~\ref{sec: Staggered electric dipole ordering} and another is the uniform electric quadrupole ordering in Sec.~\ref{sec: Uniform electric quadrupole ordering}. 
We also show the behavior of the transverse magnetization under both orderings. 

\subsection{Staggered electric dipole ordering}
\label{sec: Staggered electric dipole ordering}

\begin{figure}[tb!]
\begin{center}
\includegraphics[width=1.0\hsize]{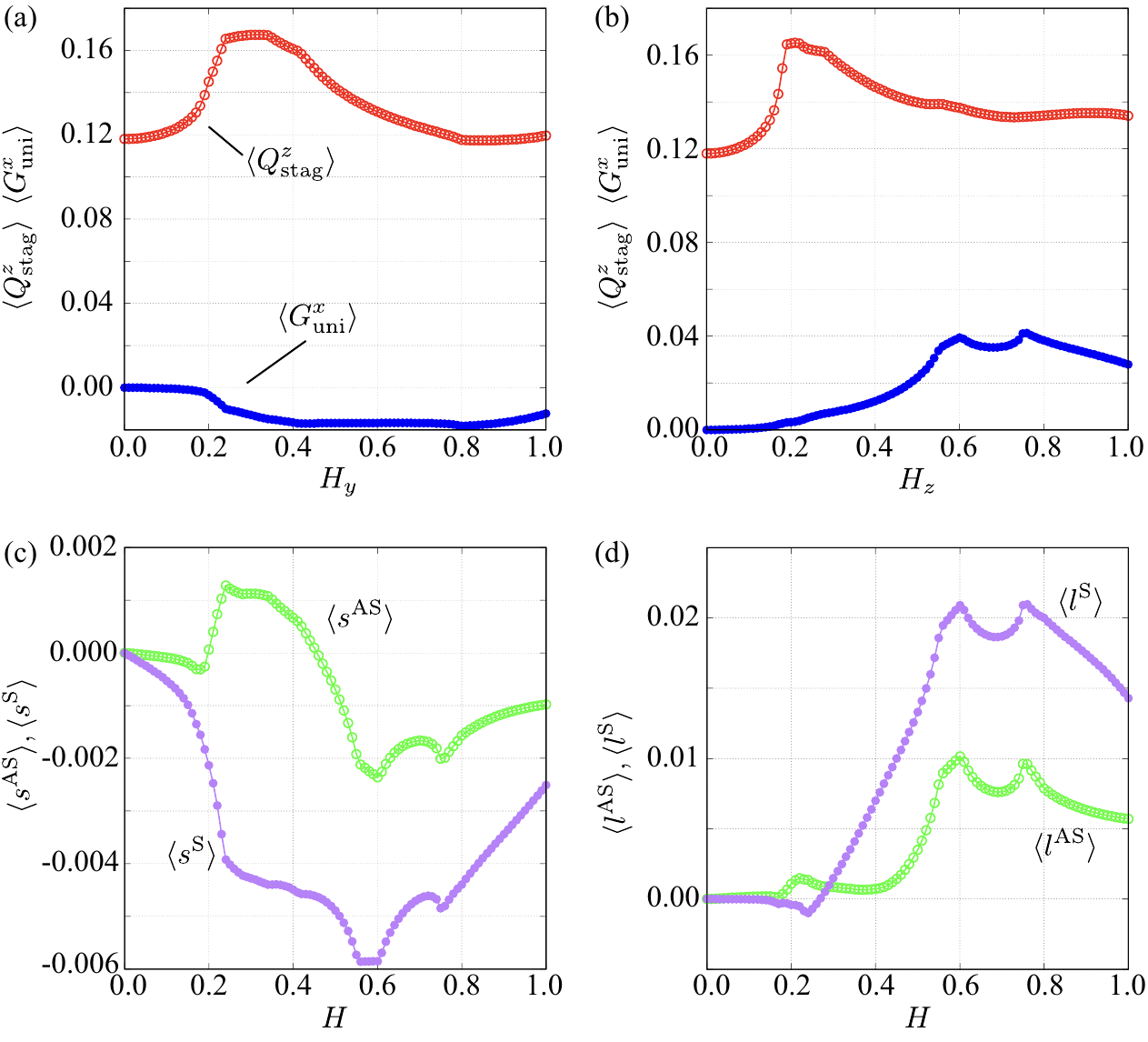} 
\caption{
\label{fig: TMag} 
(a) Magnetic field $H_y$ and (b) $H_z$ dependence of the atomic-scale electric axial moment $\langle G^x_{\rm uni}\rangle$ and cluster electric axial moment $\langle Q^z_{\rm stag} \rangle$ in the staggered electric dipole ordering for $h=0.1$ and $n_{\rm e}=2.2$. 
(c) $H$ dependence of symmetrized spin moment $\langle s^{\rm S} \rangle$ and antisymmetrized spin moment $\langle s^{\rm AS} \rangle$ for $h=0.1$ and $n_{\rm e}=2.2$. 
(d) $H$ dependence of symmetrized orbital moment $\langle l^{\rm S} \rangle$ and antisymmetrized orbital moment $\langle l^{\rm AS} \rangle$ for $h=0.1$ and $n_{\rm e}=2.2$. 
}
\end{center}
\end{figure}

As discussed in Sec.~\ref{sec: Ferroaxial moment under locally asymmetric crystal field}, the $z$ component of the staggered electric dipole ordering, $Q^z_{\rm stag} \equiv (Q^z_{\rm A}-Q^z_{\rm B})/2$ in Eq.~(\ref{eq: Qz}), in the zigzag chain accompanies the uniform electric axial moment along the $x$ direction corresponding to $G^x_{\rm uni}\equiv (G^x_{\rm A}+G^x_{\rm B})/2$ in Eq.~(\ref{eq: Gx}). 
The former corresponds to the cluster electric axial moment, while the latter corresponds to the atomic-scale electric axial moment. 
We here investigate the relationship between $Q^z_{\rm stag}$ and $G^x_{\rm uni}$ by introducing the molecular field to induce $Q^z_{\rm stag}$, which is given by 
\begin{align}
\mathcal{H}^{\rm MF}_{\rm stag}= -h \sum_{i \sigma} p_i (c^{\dagger}_{i s \sigma} c_{i p_z \sigma}+ {\rm H.c.}), 
\end{align}
where $h$ is the amplitude of the molecular field; we set $h=0.1$ in the following. 
In the presence of $\mathcal{H}^{\rm MF}_{\rm stag}$, $\langle Q^z_{\rm stag} \rangle$ ($\langle \cdots \rangle$ means the expectation value) becomes nonzero and the symmetry of the system is lowered from $D_{\rm 2h}$ to $C_{\rm 2h}$ so that the uniform electric axial moment belongs to the totally symmetric irreducible representation. 

In addition, we consider the effect of the external magnetic field in the presence of the uniform electric axial moment, since an unconventional transverse response is expected under the ordering~\cite{inda2023nonlinear}.
Then, we introduce the Zeeman Hamiltonian given by 
\begin{align} 
\label{eq: HamZeeman}
\mathcal{H}^{\rm Z}&=-\sum_{i} \bm{H} \cdot  (\bm{l}_i + 2 \bm{s}_i), 
\end{align}
where we consider the transverse magnetic field perpendicular to the uniform electric axial moment, i.e., $\bm{H}=(0, H_y, H_z)$. 
When the uniform electric axial moment is activated, the $y$ ($z$) component of the uniform magnetization, $M^{y}$ $(M^z)$, is expected to be induced by the magnetic field along the $z$ ($y$) direction in addition to the parallel component; we denote the notation to represent such a situation as $M^y (H_z)$ [$M^z (H_y)$].
In particular, the antisymmetrized component $M^{\rm AS} \equiv [M^y(H_z)-M^z(H_y)]/2$ corresponds to the signal of the electric axial moment, while the symmetrized component $M^{\rm S} \equiv [M^y(H_z)+M^z(H_y)]/2$ corresponds to that of the $yz$ component of the electric quadrupole moment. 
In the present model, both components are permitted from the $C_{\rm 2h}$ symmetry, since the staggered electric dipole ordering is regarded as the superposition of the electric toroidal dipole and electric quadrupole~\cite{hayami2016emergent}. 
We calculate the antisymmetrized (symmetrized) spin and orbital moments $s^{\rm AS}$ and $l^{\rm AS}$ ($s^{\rm S}$ and $l^{\rm S}$), respectively; $M^{\rm AS}=s^{\rm AS}+l^{\rm AS}$ and $M^{\rm S}=s^{\rm S}+l^{\rm S}$. 

Figures~\ref{fig: TMag}(a) and \ref{fig: TMag}(b) show the $H_y$ and $H_z$ dependence of the expectation values of cluster and atomic electric axial moment, $\langle Q^z_{\rm stag} \rangle$ and $\langle G^x_{\rm uni} \rangle$, respectively. 
We set the electron filling as $n_{\rm e}=2.2$, where $n_{\rm e}=8$ is the full filling. 
In both cases, $\langle Q^z_{\rm stag} \rangle$ becomes nonzero for $H_y \geq 0$ or $H_z \geq 0$, while $\langle G^x_{\rm uni} \rangle$ becomes nonzero in a nonzero field for $H_y > 0$ or $H_z > 0$. 
This indicates that $\langle Q^z_{\rm stag} \rangle$ and $\langle G^x_{\rm uni} \rangle$ are related to each other via the external magnetic field. 
Indeed, we obtain the essential model parameters under $H_z$ in the form of $h H_z^2 (c_1 V + c_2 t_{sp\sigma} ) F $, where $c_1$ and $c_2$ are numerical coefficients and $F$ denotes the function depending on the model parameters of the Hamiltonian. 
One finds that $\langle G^x_{\rm uni} \rangle$ can be induced in the same direction irrespective of the sign of $H_z$~\cite{hayami2023planar} and needs the odd-parity $s$-$p$ hybridization ($V$ and $t_{sp\sigma}$) as well as the molecular field ($h$).
A similar relation also holds for the case under $H_y$ in Fig.~\ref{fig: TMag}(a). 

Next, we plot the expectation values of $s^{\rm AS}$ and $s^{\rm S}$ in Fig.~\ref{fig: TMag}(c) and $l^{\rm AS}$ and $l^{\rm S}$ in Fig.~\ref{fig: TMag}(d) with $H_y=H_z=H$. 
The result shows that both quantities become nonzero once the magnetic field is applied. 
By analyzing the essential model parameters for $\langle s^{\rm AS} \rangle$, $\langle s^{\rm S} \rangle$, $\langle l^{\rm AS} \rangle$, and $\langle l^{\rm S} \rangle$, we find a similar tendency of the model parameter dependence to induce $K_{z;x}$ and $\langle G^x_{\rm uni} \rangle$; $\langle s^{\rm AS} \rangle$ and $\langle s^{\rm S} \rangle $ are proportional to $h  H \lambda (c'_1 V + c'_2 t_{sp\sigma} ) F' $, while $\langle l^{\rm AS} \rangle$ and $\langle l^{\rm S} \rangle$ are proportional to $h H  (c''_1 V + c''_2 t_{sp\sigma} ) F'' $, where $c'_1$, $c'_2$, $c''_1$, and $c''_2$ are numerical coefficients and $F'$ and $F''$ are functions of the model parameters~\cite{comment_essential_sl}. 
The difference is that the spin--orbit coupling $\lambda$ is included in the essential model parameters of $\langle s^{\rm AS}\rangle $ and $\langle s^{\rm S}\rangle$, since the odd-parity hybridization without spin dependence affects $\langle s^{\rm AS} \rangle $ and $\langle s^{\rm S} \rangle $ via the spin--orbit coupling. 
Another difference from $\langle G^x_{\rm uni} \rangle$ appears in the magnetic field dependence. 
The transverse magnetization is proportional to $H$ rather than $H^2$ owing to the different time-reversal parity between $\langle G^x_{\rm uni} \rangle$ and spin/orbital moments.

\subsection{Uniform electric quadrupole ordering}
\label{sec: Uniform electric quadrupole ordering}

\begin{figure}[tb!]
\begin{center}
\includegraphics[width=1.0\hsize]{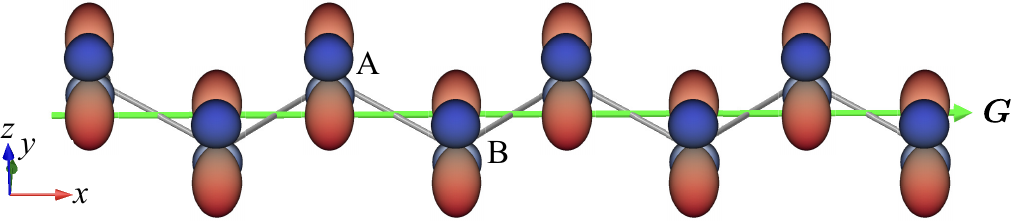} 
\caption{
\label{fig: ponti_Qyz} 
Uniform alignment of electric quadrupole $Q_{yz}$ in the zigzag chain. 
The uniform electric axial moment denoted by the green arrow is induced along the $x$ direction. 
}
\end{center}
\end{figure}

\begin{figure}[tb!]
\begin{center}
\includegraphics[width=1.0\hsize]{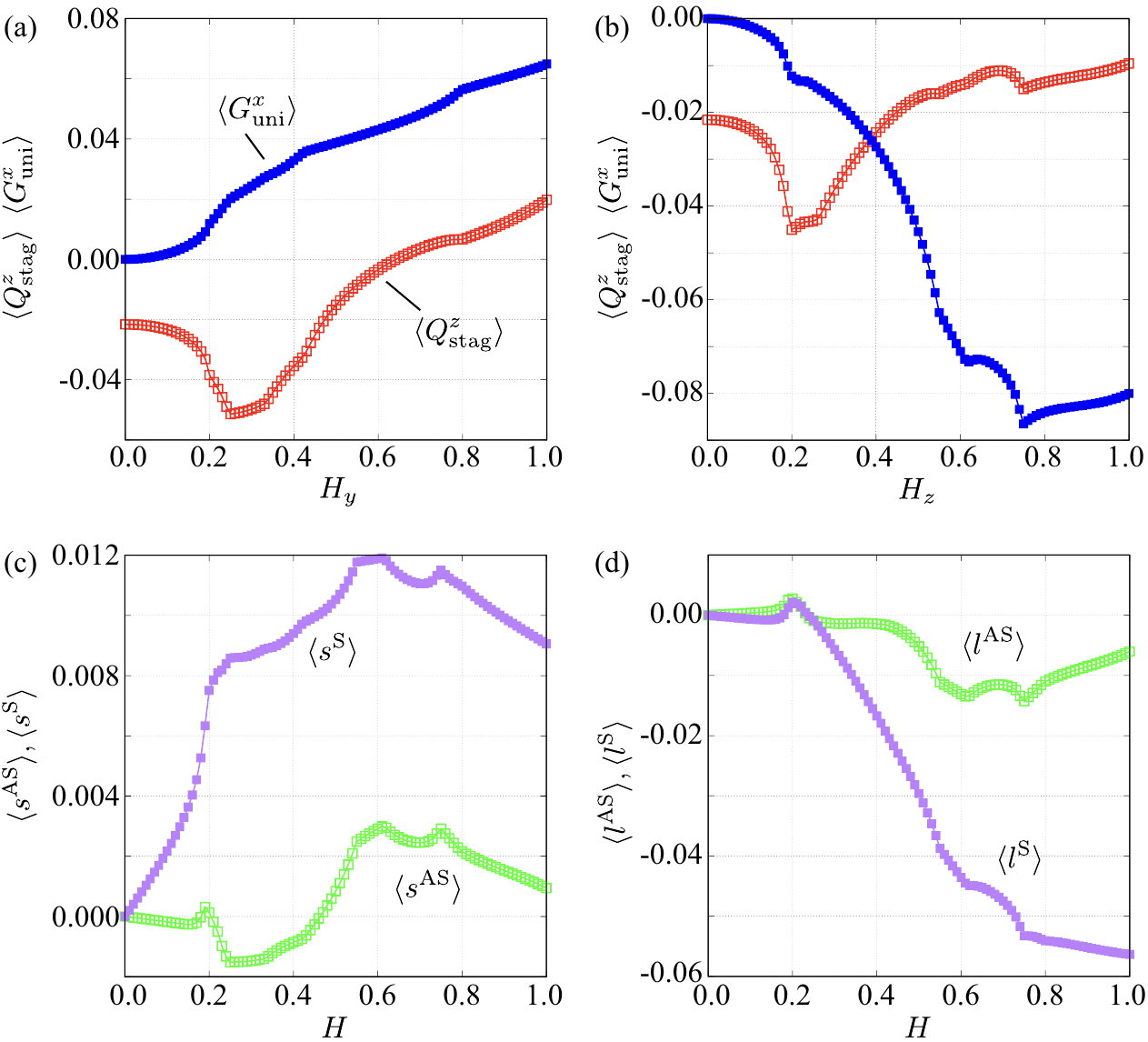} 
\caption{
\label{fig: TMag_Qyz} 
(a) $H_y$ and (b) $H_z$ dependence of $\langle G^x_{\rm uni}\rangle$ and $\langle Q^z_{\rm stag} \rangle$ in the uniform electric quadrupole ordering for $h'=0.1$ and $n_{\rm e}=2.2$. 
(c) $H$ dependence of $\langle s^{\rm S} \rangle$ and $\langle s^{\rm AS} \rangle$ for $h'=0.1$ and $n_{\rm e}=2.2$. 
(d) $H$ dependence of $\langle l^{\rm S} \rangle$ and $\langle l^{\rm AS} \rangle$ for $h'=0.1$ and $n_{\rm e}=2.2$. 
}
\end{center}
\end{figure}

We have so far investigated the situation where the uniform electric axial moment is induced by the staggered electric dipole, as schematically shown in Fig.~\ref{fig: ponti}(d). 
Meanwhile, the uniform electric axial moment is expected to be induced even for other electronic orderings once the resultant symmetry in the presence of the symmetry-lowering molecular field is the same as that in the case of the staggered electric dipole ordering. 
As an example, we consider the uniform electric quadrupole ordering, as shown in Fig.~\ref{fig: ponti_Qyz}, which also lowers the symmetry from $D_{\rm 2h}$ to $C_{\rm 2h}$. 
The microscopic order parameter is given by  
\begin{align}
\mathcal{H}^{\rm MF}_{\rm uni}= -h' \sum_{i \sigma} (c^{\dagger}_{i p_y \sigma} c_{i p_z \sigma}+ {\rm H.c.}), 
\end{align}
where $h'$ is the magnitude of the molecular field regarding the electric quadrupole. 

Figures~\ref{fig: TMag_Qyz}(a) and \ref{fig: TMag_Qyz}(b) show the $H_y$ and $H_z$ dependence of $\langle Q^z_{\rm stag} \rangle$ and $\langle G^x_{\rm uni} \rangle$ for the model Hamiltonian $\mathcal{H}+\mathcal{H}^{\rm MF}_{\rm uni}+\mathcal{H}^{\rm Z}$. 
We also show the $H$ dependence of $\langle s^{\rm AS} \rangle$ and $\langle s^{\rm S} \rangle$ ($\langle l^{\rm AS} \rangle$ and $\langle l^{\rm S} \rangle$) in Fig.~\ref{fig: TMag_Qyz}(c) [Fig.~\ref{fig: TMag_Qyz}(d)]. 
In contrast to the result in Fig.~\ref{fig: TMag}, $\langle G^x_{\rm uni} \rangle$, $\langle s^{\rm AS} \rangle$ and $\langle s^{\rm S} \rangle$, $\langle l^{\rm AS} \rangle$, and $\langle l^{\rm S} \rangle$ exhibit different model parameter dependence: $\langle G^x_{\rm uni} \rangle$ is proportional to $h' H^2 F'''$, $\langle s^{\rm AS} \rangle$ and $\langle s^{\rm S} \rangle $ are proportional to $h' H \lambda F'''' $, and $\langle l^{\rm AS} \rangle $ and $\langle l^{\rm S} \rangle $ are proportional to $h' H F''''' $, where $F'''$, $F''''$, $F'''''$ are functions of the model parameters~\cite{comment_essential_sl}. 
Thus, the effect of the odd-parity hybridization, $V$ and $t_{sp\sigma}$, is not required in the case of the uniform electric quadrupole ordering. 
The difference is understood from the orbital dependence of the order parameter; the electric dipole ($Q^z$) corresponds to the hybridization between the $s$ and $p$ orbitals, while the electric quadrupole ($Q_{yz}$) corresponds to the hybridization between the $p$ orbitals. 
Since the operator of the electric axial moment $G^x$ is described by the hybridization between $p$ orbitals in Eq.~(\ref{eq: Gx}), the $s$-$p$ hybridized factor is required in order to couple $Q^z$ and $G^x$ in the case of the staggered electric dipole ordering, while it is not necessary for the coupling between $Q_{yz}$ and $G^x$ in the case of the uniform electric quadrupole ordering. 
In this way, the important model parameters to induce the electric axial moment and its related transverse magnetization largely depend on the microscopic electronic order parameters.

\section{Summary}
\label{sec: Summary}

To summarize, we have investigated the fundamental properties of the electric axial moment by focusing on its ordering in the one-dimensional zigzag chain. 
We proposed that the zigzag chain with a locally noncentrosymmetric lattice structure is a prototype to examine electric-axial physics by exemplifying the emergence of staggered and uniform electric axial moments.  
We found that the staggered electric axial moment is induced by applying the external electric field, where the interplay between the spin--orbit coupling and the odd-parity hybridization plays an important role. 
In addition, we have shown that the uniform electric axial moment is activated by the staggered electric dipole ordering and/or uniform electric quadrupole ordering, both of which are a source of the transverse magnetic responses. 
Since there are a lot of materials with the zigzag structure, such as LnM$_2$Al$_{10}$ (Ln: lanthanoid ions, M: transition metal ions)~\cite{thiede1998ternary, reehuis2003magnetic, KhalyavinPhysRevB.82.100405, TanidaPhysRevB.84.115128, mignot2011neutron, muro2011magnetic, kato2011magnetic} and U$_2$Ir$_3$Si$_5$~\cite{Li_PhysRevB.99.054408}, our results will be a reference to study the nature of the electric axial moment in future experiments.

\begin{acknowledgments}
This research was supported by JSPS KAKENHI Grants Numbers JP21H01037, JP22H04468, JP22H00101, JP22H01183, JP23K03288, JP23H04869, and by JST PRESTO (JPMJPR20L8). 
Parts of the numerical calculations were performed in the supercomputing systems in ISSP, the University of Tokyo.
\end{acknowledgments}

%\appendix

\bibliographystyle{apsrev}
\bibliography{../ref}

\begin{thebibliography}{135}
\expandafter\ifx\csname natexlab\endcsname\relax\def\natexlab#1{#1}\fi
\expandafter\ifx\csname bibnamefont\endcsname\relax
  \def\bibnamefont#1{#1}\fi
\expandafter\ifx\csname bibfnamefont\endcsname\relax
  \def\bibfnamefont#1{#1}\fi
\expandafter\ifx\csname citenamefont\endcsname\relax
  \def\citenamefont#1{#1}\fi
\expandafter\ifx\csname url\endcsname\relax
  \def\url#1{\texttt{#1}}\fi
\expandafter\ifx\csname urlprefix\endcsname\relax\def\urlprefix{URL }\fi
\providecommand{\bibinfo}[2]{#2}
\providecommand{\eprint}[2][]{\url{#2}}

\bibitem[{\citenamefont{Fu}(2015)}]{Fu_PhysRevLett.115.026401}
\bibinfo{author}{\bibfnamefont{L.}~\bibnamefont{Fu}}, \bibinfo{journal}{Phys.
  Rev. Lett.} \textbf{\bibinfo{volume}{115}}, \bibinfo{pages}{026401}
  (\bibinfo{year}{2015}).

\bibitem[{\citenamefont{Kozii and Fu}(2015)}]{Kozii_PhysRevLett.115.207002}
\bibinfo{author}{\bibfnamefont{V.}~\bibnamefont{Kozii}} \bibnamefont{and}
  \bibinfo{author}{\bibfnamefont{L.}~\bibnamefont{Fu}}, \bibinfo{journal}{Phys.
  Rev. Lett.} \textbf{\bibinfo{volume}{115}}, \bibinfo{pages}{207002}
  (\bibinfo{year}{2015}).

\bibitem[{\citenamefont{Venderbos et~al.}(2016)\citenamefont{Venderbos, Kozii,
  and Fu}}]{Venderbos_PhysRevB.94.180504}
\bibinfo{author}{\bibfnamefont{J.~W.~F.} \bibnamefont{Venderbos}},
  \bibinfo{author}{\bibfnamefont{V.}~\bibnamefont{Kozii}}, \bibnamefont{and}
  \bibinfo{author}{\bibfnamefont{L.}~\bibnamefont{Fu}}, \bibinfo{journal}{Phys.
  Rev. B} \textbf{\bibinfo{volume}{94}}, \bibinfo{pages}{180504(R)}
  (\bibinfo{year}{2016}).

\bibitem[{\citenamefont{Rashba}(1960)}]{rashba1960properties}
\bibinfo{author}{\bibfnamefont{E.~I.} \bibnamefont{Rashba}},
  \bibinfo{journal}{Sov. Phys. Solid State} \textbf{\bibinfo{volume}{2}},
  \bibinfo{pages}{1109} (\bibinfo{year}{1960}).

\bibitem[{\citenamefont{Dresselhaus et~al.}(2008)\citenamefont{Dresselhaus,
  Dresselhaus, and Jorio}}]{Dresselhaus_Dresselhaus_Jorio}
\bibinfo{author}{\bibfnamefont{M.~S.} \bibnamefont{Dresselhaus}},
  \bibinfo{author}{\bibfnamefont{G.}~\bibnamefont{Dresselhaus}},
  \bibnamefont{and} \bibinfo{author}{\bibfnamefont{A.}~\bibnamefont{Jorio}},
  \emph{\bibinfo{title}{Group Theory: Application to the Physics of Condensed
  Matter}} (\bibinfo{publisher}{Springer-Verlag}, \bibinfo{address}{Berlin
  Heidelberg}, \bibinfo{year}{2008}).

\bibitem[{\citenamefont{Murakami et~al.}(2003)\citenamefont{Murakami, Nagaosa,
  and Zhang}}]{murakami2003dissipationless}
\bibinfo{author}{\bibfnamefont{S.}~\bibnamefont{Murakami}},
  \bibinfo{author}{\bibfnamefont{N.}~\bibnamefont{Nagaosa}}, \bibnamefont{and}
  \bibinfo{author}{\bibfnamefont{S.-C.} \bibnamefont{Zhang}},
  \bibinfo{journal}{Science} \textbf{\bibinfo{volume}{301}},
  \bibinfo{pages}{1348} (\bibinfo{year}{2003}).

\bibitem[{\citenamefont{Murakami et~al.}(2004)\citenamefont{Murakami, Nagaosa,
  and Zhang}}]{Murakami_PhysRevLett.93.156804}
\bibinfo{author}{\bibfnamefont{S.}~\bibnamefont{Murakami}},
  \bibinfo{author}{\bibfnamefont{N.}~\bibnamefont{Nagaosa}}, \bibnamefont{and}
  \bibinfo{author}{\bibfnamefont{S.-C.} \bibnamefont{Zhang}},
  \bibinfo{journal}{Phys. Rev. Lett.} \textbf{\bibinfo{volume}{93}},
  \bibinfo{pages}{156804} (\bibinfo{year}{2004}).

\bibitem[{\citenamefont{Sinova et~al.}(2004)\citenamefont{Sinova, Culcer, Niu,
  Sinitsyn, Jungwirth, and MacDonald}}]{Sinova_PhysRevLett.92.126603}
\bibinfo{author}{\bibfnamefont{J.}~\bibnamefont{Sinova}},
  \bibinfo{author}{\bibfnamefont{D.}~\bibnamefont{Culcer}},
  \bibinfo{author}{\bibfnamefont{Q.}~\bibnamefont{Niu}},
  \bibinfo{author}{\bibfnamefont{N.~A.} \bibnamefont{Sinitsyn}},
  \bibinfo{author}{\bibfnamefont{T.}~\bibnamefont{Jungwirth}},
  \bibnamefont{and} \bibinfo{author}{\bibfnamefont{A.~H.}
  \bibnamefont{MacDonald}}, \bibinfo{journal}{Phys. Rev. Lett.}
  \textbf{\bibinfo{volume}{92}}, \bibinfo{pages}{126603}
  (\bibinfo{year}{2004}).

\bibitem[{\citenamefont{Fujimoto}(2006)}]{Fujimoto_doi:10.1143/JPSJ.75.083704}
\bibinfo{author}{\bibfnamefont{S.}~\bibnamefont{Fujimoto}},
  \bibinfo{journal}{J. Phys. Soc. Jpn.} \textbf{\bibinfo{volume}{75}},
  \bibinfo{pages}{083704} (\bibinfo{year}{2006}).

\bibitem[{\citenamefont{Fujimoto}(2007)}]{fujimoto2007fermi}
\bibinfo{author}{\bibfnamefont{S.}~\bibnamefont{Fujimoto}},
  \bibinfo{journal}{J. Phys. Soc. Jpn.} \textbf{\bibinfo{volume}{76}}
  (\bibinfo{year}{2007}).

\bibitem[{\citenamefont{Edelstein}(1990)}]{edelstein1990spin}
\bibinfo{author}{\bibfnamefont{V.~M.} \bibnamefont{Edelstein}},
  \bibinfo{journal}{Solid State Commun.} \textbf{\bibinfo{volume}{73}},
  \bibinfo{pages}{233} (\bibinfo{year}{1990}).

\bibitem[{\citenamefont{Yip}(2002)}]{Yip_PhysRevB.65.144508}
\bibinfo{author}{\bibfnamefont{S.~K.} \bibnamefont{Yip}},
  \bibinfo{journal}{Phys. Rev. B} \textbf{\bibinfo{volume}{65}},
  \bibinfo{pages}{144508} (\bibinfo{year}{2002}).

\bibitem[{\citenamefont{Fujimoto}(2005)}]{Fujimoto_PhysRevB.72.024515}
\bibinfo{author}{\bibfnamefont{S.}~\bibnamefont{Fujimoto}},
  \bibinfo{journal}{Phys. Rev. B} \textbf{\bibinfo{volume}{72}},
  \bibinfo{pages}{024515} (\bibinfo{year}{2005}).

\bibitem[{\citenamefont{Yoda et~al.}(2018)\citenamefont{Yoda, Yokoyama, and
  Murakami}}]{yoda2018orbital}
\bibinfo{author}{\bibfnamefont{T.}~\bibnamefont{Yoda}},
  \bibinfo{author}{\bibfnamefont{T.}~\bibnamefont{Yokoyama}}, \bibnamefont{and}
  \bibinfo{author}{\bibfnamefont{S.}~\bibnamefont{Murakami}},
  \bibinfo{journal}{Nano letters} \textbf{\bibinfo{volume}{18}},
  \bibinfo{pages}{916} (\bibinfo{year}{2018}).

\bibitem[{\citenamefont{Massarelli et~al.}(2019)\citenamefont{Massarelli, Wu,
  and Paramekanti}}]{Massarelli_PhysRevB.100.075136}
\bibinfo{author}{\bibfnamefont{G.}~\bibnamefont{Massarelli}},
  \bibinfo{author}{\bibfnamefont{B.}~\bibnamefont{Wu}}, \bibnamefont{and}
  \bibinfo{author}{\bibfnamefont{A.}~\bibnamefont{Paramekanti}},
  \bibinfo{journal}{Phys. Rev. B} \textbf{\bibinfo{volume}{100}},
  \bibinfo{pages}{075136} (\bibinfo{year}{2019}).

\bibitem[{\citenamefont{Sodemann and
  Fu}(2015)}]{Sodemann_PhysRevLett.115.216806}
\bibinfo{author}{\bibfnamefont{I.}~\bibnamefont{Sodemann}} \bibnamefont{and}
  \bibinfo{author}{\bibfnamefont{L.}~\bibnamefont{Fu}}, \bibinfo{journal}{Phys.
  Rev. Lett.} \textbf{\bibinfo{volume}{115}}, \bibinfo{pages}{216806}
  (\bibinfo{year}{2015}).

\bibitem[{\citenamefont{Nandy and Sodemann}(2019)}]{Nandy_PhysRevB.100.195117}
\bibinfo{author}{\bibfnamefont{S.}~\bibnamefont{Nandy}} \bibnamefont{and}
  \bibinfo{author}{\bibfnamefont{I.}~\bibnamefont{Sodemann}},
  \bibinfo{journal}{Phys. Rev. B} \textbf{\bibinfo{volume}{100}},
  \bibinfo{pages}{195117} (\bibinfo{year}{2019}).

\bibitem[{\citenamefont{Dubovik and Cheshkov}(1975)}]{dubovik1975multipole}
\bibinfo{author}{\bibfnamefont{V.}~\bibnamefont{Dubovik}} \bibnamefont{and}
  \bibinfo{author}{\bibfnamefont{A.}~\bibnamefont{Cheshkov}},
  \bibinfo{journal}{Sov. J. Part. Nucl} \textbf{\bibinfo{volume}{5}},
  \bibinfo{pages}{318} (\bibinfo{year}{1975}).

\bibitem[{\citenamefont{Dubovik and Tugushev}(1990)}]{dubovik1990toroid}
\bibinfo{author}{\bibfnamefont{V.}~\bibnamefont{Dubovik}} \bibnamefont{and}
  \bibinfo{author}{\bibfnamefont{V.}~\bibnamefont{Tugushev}},
  \bibinfo{journal}{Phys. Rep.} \textbf{\bibinfo{volume}{187}},
  \bibinfo{pages}{145} (\bibinfo{year}{1990}).

\bibitem[{\citenamefont{Gorbatsevich and
  Kopaev}(1994)}]{gorbatsevich1994toroidal}
\bibinfo{author}{\bibfnamefont{A.}~\bibnamefont{Gorbatsevich}}
  \bibnamefont{and} \bibinfo{author}{\bibfnamefont{Y.~V.}
  \bibnamefont{Kopaev}}, \bibinfo{journal}{Ferroelectrics}
  \textbf{\bibinfo{volume}{161}}, \bibinfo{pages}{321} (\bibinfo{year}{1994}).

\bibitem[{\citenamefont{Spaldin et~al.}(2008)\citenamefont{Spaldin, Fiebig, and
  Mostovoy}}]{Spaldin_0953-8984-20-43-434203}
\bibinfo{author}{\bibfnamefont{N.~A.} \bibnamefont{Spaldin}},
  \bibinfo{author}{\bibfnamefont{M.}~\bibnamefont{Fiebig}}, \bibnamefont{and}
  \bibinfo{author}{\bibfnamefont{M.}~\bibnamefont{Mostovoy}},
  \bibinfo{journal}{J. Phys.: Condens. Matter} \textbf{\bibinfo{volume}{20}},
  \bibinfo{pages}{434203} (\bibinfo{year}{2008}).

\bibitem[{\citenamefont{Van~Aken et~al.}(2007)\citenamefont{Van~Aken, Rivera,
  Schmid, and Fiebig}}]{van2007observation}
\bibinfo{author}{\bibfnamefont{B.~B.} \bibnamefont{Van~Aken}},
  \bibinfo{author}{\bibfnamefont{J.-P.} \bibnamefont{Rivera}},
  \bibinfo{author}{\bibfnamefont{H.}~\bibnamefont{Schmid}}, \bibnamefont{and}
  \bibinfo{author}{\bibfnamefont{M.}~\bibnamefont{Fiebig}},
  \bibinfo{journal}{Nature} \textbf{\bibinfo{volume}{449}},
  \bibinfo{pages}{702} (\bibinfo{year}{2007}).

\bibitem[{\citenamefont{Cheong et~al.}(2018)\citenamefont{Cheong, Talbayev,
  Kiryukhin, and Saxena}}]{cheong2018broken}
\bibinfo{author}{\bibfnamefont{S.-W.} \bibnamefont{Cheong}},
  \bibinfo{author}{\bibfnamefont{D.}~\bibnamefont{Talbayev}},
  \bibinfo{author}{\bibfnamefont{V.}~\bibnamefont{Kiryukhin}},
  \bibnamefont{and} \bibinfo{author}{\bibfnamefont{A.}~\bibnamefont{Saxena}},
  \bibinfo{journal}{npj Quantum Mater.} \textbf{\bibinfo{volume}{3}},
  \bibinfo{pages}{19} (\bibinfo{year}{2018}).

\bibitem[{\citenamefont{Popov et~al.}(1999)\citenamefont{Popov, Kadomtseva,
  Belov, Vorob'ev, and Zvezdin}}]{popov1999magnetic}
\bibinfo{author}{\bibfnamefont{Y.~F.} \bibnamefont{Popov}},
  \bibinfo{author}{\bibfnamefont{A.}~\bibnamefont{Kadomtseva}},
  \bibinfo{author}{\bibfnamefont{D.}~\bibnamefont{Belov}},
  \bibinfo{author}{\bibfnamefont{G.}~\bibnamefont{Vorob'ev}}, \bibnamefont{and}
  \bibinfo{author}{\bibfnamefont{A.}~\bibnamefont{Zvezdin}},
  \bibinfo{journal}{J. Exp. Theor. Phys. Lett.} \textbf{\bibinfo{volume}{69}},
  \bibinfo{pages}{330} (\bibinfo{year}{1999}).

\bibitem[{\citenamefont{Schmid}(2001)}]{schmid2001ferrotoroidics}
\bibinfo{author}{\bibfnamefont{H.}~\bibnamefont{Schmid}},
  \bibinfo{journal}{Ferroelectrics} \textbf{\bibinfo{volume}{252}},
  \bibinfo{pages}{41} (\bibinfo{year}{2001}).

\bibitem[{\citenamefont{Ederer and Spaldin}(2007)}]{EdererPhysRevB.76.214404}
\bibinfo{author}{\bibfnamefont{C.}~\bibnamefont{Ederer}} \bibnamefont{and}
  \bibinfo{author}{\bibfnamefont{N.~A.} \bibnamefont{Spaldin}},
  \bibinfo{journal}{Phys. Rev. B} \textbf{\bibinfo{volume}{76}},
  \bibinfo{pages}{214404} (\bibinfo{year}{2007}).

\bibitem[{\citenamefont{Khomskii}(2009)}]{KhomskiiPhysics.2.20}
\bibinfo{author}{\bibfnamefont{D.}~\bibnamefont{Khomskii}},
  \bibinfo{journal}{Physics} \textbf{\bibinfo{volume}{2}}, \bibinfo{pages}{20}
  (\bibinfo{year}{2009}).

\bibitem[{\citenamefont{Zimmermann et~al.}(2014)\citenamefont{Zimmermann,
  Meier, and Fiebig}}]{zimmermann2014ferroic}
\bibinfo{author}{\bibfnamefont{A.~S.} \bibnamefont{Zimmermann}},
  \bibinfo{author}{\bibfnamefont{D.}~\bibnamefont{Meier}}, \bibnamefont{and}
  \bibinfo{author}{\bibfnamefont{M.}~\bibnamefont{Fiebig}},
  \bibinfo{journal}{Nat. Commun.} \textbf{\bibinfo{volume}{5}},
  \bibinfo{pages}{4796} (\bibinfo{year}{2014}).

\bibitem[{\citenamefont{Sawada and
  Nagaosa}(2005)}]{Sawada_PhysRevLett.95.237402}
\bibinfo{author}{\bibfnamefont{K.}~\bibnamefont{Sawada}} \bibnamefont{and}
  \bibinfo{author}{\bibfnamefont{N.}~\bibnamefont{Nagaosa}},
  \bibinfo{journal}{Phys. Rev. Lett.} \textbf{\bibinfo{volume}{95}},
  \bibinfo{pages}{237402} (\bibinfo{year}{2005}).

\bibitem[{\citenamefont{Miyahara and Furukawa}(2012)}]{Miyahara_JPSJ.81.023712}
\bibinfo{author}{\bibfnamefont{S.}~\bibnamefont{Miyahara}} \bibnamefont{and}
  \bibinfo{author}{\bibfnamefont{N.}~\bibnamefont{Furukawa}},
  \bibinfo{journal}{J. Phys. Soc. Jpn.} \textbf{\bibinfo{volume}{81}},
  \bibinfo{pages}{023712} (\bibinfo{year}{2012}).

\bibitem[{\citenamefont{Miyahara and
  Furukawa}(2014)}]{Miyahara_PhysRevB.89.195145}
\bibinfo{author}{\bibfnamefont{S.}~\bibnamefont{Miyahara}} \bibnamefont{and}
  \bibinfo{author}{\bibfnamefont{N.}~\bibnamefont{Furukawa}},
  \bibinfo{journal}{Phys. Rev. B} \textbf{\bibinfo{volume}{89}},
  \bibinfo{pages}{195145} (\bibinfo{year}{2014}).

\bibitem[{\citenamefont{Tokura and Nagaosa}(2018)}]{tokura2018nonreciprocal}
\bibinfo{author}{\bibfnamefont{Y.}~\bibnamefont{Tokura}} \bibnamefont{and}
  \bibinfo{author}{\bibfnamefont{N.}~\bibnamefont{Nagaosa}},
  \bibinfo{journal}{Nat. Commun.} \textbf{\bibinfo{volume}{9}},
  \bibinfo{pages}{3740} (\bibinfo{year}{2018}).

\bibitem[{\citenamefont{Ye et~al.}(1999)\citenamefont{Ye, Kim, Millis,
  Shraiman, Majumdar, and Te\ifmmode \check{s}\else
  \v{s}\fi{}anovi\ifmmode~\acute{c}\else \'{c}\fi{}}}]{Ye_PhysRevLett.83.3737}
\bibinfo{author}{\bibfnamefont{J.}~\bibnamefont{Ye}},
  \bibinfo{author}{\bibfnamefont{Y.~B.} \bibnamefont{Kim}},
  \bibinfo{author}{\bibfnamefont{A.~J.} \bibnamefont{Millis}},
  \bibinfo{author}{\bibfnamefont{B.~I.} \bibnamefont{Shraiman}},
  \bibinfo{author}{\bibfnamefont{P.}~\bibnamefont{Majumdar}}, \bibnamefont{and}
  \bibinfo{author}{\bibfnamefont{Z.}~\bibnamefont{Te\ifmmode \check{s}\else
  \v{s}\fi{}anovi\ifmmode~\acute{c}\else \'{c}\fi{}}}, \bibinfo{journal}{Phys.
  Rev. Lett.} \textbf{\bibinfo{volume}{83}}, \bibinfo{pages}{3737}
  (\bibinfo{year}{1999}).

\bibitem[{\citenamefont{Solovyev}(1997)}]{Solovyev_PhysRevB.55.8060}
\bibinfo{author}{\bibfnamefont{I.~V.} \bibnamefont{Solovyev}},
  \bibinfo{journal}{Phys. Rev. B} \textbf{\bibinfo{volume}{55}},
  \bibinfo{pages}{8060} (\bibinfo{year}{1997}).

\bibitem[{\citenamefont{Chen}(2022)}]{Chen_PhysRevB.106.024421}
\bibinfo{author}{\bibfnamefont{H.}~\bibnamefont{Chen}}, \bibinfo{journal}{Phys.
  Rev. B} \textbf{\bibinfo{volume}{106}}, \bibinfo{pages}{024421}
  (\bibinfo{year}{2022}).

\bibitem[{\citenamefont{Naka et~al.}(2020)\citenamefont{Naka, Hayami, Kusunose,
  Yanagi, Motome, and Seo}}]{Naka_PhysRevB.102.075112}
\bibinfo{author}{\bibfnamefont{M.}~\bibnamefont{Naka}},
  \bibinfo{author}{\bibfnamefont{S.}~\bibnamefont{Hayami}},
  \bibinfo{author}{\bibfnamefont{H.}~\bibnamefont{Kusunose}},
  \bibinfo{author}{\bibfnamefont{Y.}~\bibnamefont{Yanagi}},
  \bibinfo{author}{\bibfnamefont{Y.}~\bibnamefont{Motome}}, \bibnamefont{and}
  \bibinfo{author}{\bibfnamefont{H.}~\bibnamefont{Seo}},
  \bibinfo{journal}{Phys. Rev. B} \textbf{\bibinfo{volume}{102}},
  \bibinfo{pages}{075112} (\bibinfo{year}{2020}).

\bibitem[{\citenamefont{Hayami and
  Kusunose}(2021{\natexlab{a}})}]{Hayami_PhysRevB.103.L180407}
\bibinfo{author}{\bibfnamefont{S.}~\bibnamefont{Hayami}} \bibnamefont{and}
  \bibinfo{author}{\bibfnamefont{H.}~\bibnamefont{Kusunose}},
  \bibinfo{journal}{Phys. Rev. B} \textbf{\bibinfo{volume}{103}},
  \bibinfo{pages}{L180407} (\bibinfo{year}{2021}{\natexlab{a}}).

\bibitem[{\citenamefont{Tomizawa and
  Kontani}(2009)}]{Tomizawa_PhysRevB.80.100401}
\bibinfo{author}{\bibfnamefont{T.}~\bibnamefont{Tomizawa}} \bibnamefont{and}
  \bibinfo{author}{\bibfnamefont{H.}~\bibnamefont{Kontani}},
  \bibinfo{journal}{Phys. Rev. B} \textbf{\bibinfo{volume}{80}},
  \bibinfo{pages}{100401} (\bibinfo{year}{2009}).

\bibitem[{\citenamefont{Chen et~al.}(2014)\citenamefont{Chen, Niu, and
  MacDonald}}]{Chen_PhysRevLett.112.017205}
\bibinfo{author}{\bibfnamefont{H.}~\bibnamefont{Chen}},
  \bibinfo{author}{\bibfnamefont{Q.}~\bibnamefont{Niu}}, \bibnamefont{and}
  \bibinfo{author}{\bibfnamefont{A.~H.} \bibnamefont{MacDonald}},
  \bibinfo{journal}{Phys. Rev. Lett.} \textbf{\bibinfo{volume}{112}},
  \bibinfo{pages}{017205} (\bibinfo{year}{2014}).

\bibitem[{\citenamefont{Nakatsuji et~al.}(2015)\citenamefont{Nakatsuji,
  Kiyohara, and Higo}}]{nakatsuji2015large}
\bibinfo{author}{\bibfnamefont{S.}~\bibnamefont{Nakatsuji}},
  \bibinfo{author}{\bibfnamefont{N.}~\bibnamefont{Kiyohara}}, \bibnamefont{and}
  \bibinfo{author}{\bibfnamefont{T.}~\bibnamefont{Higo}},
  \bibinfo{journal}{Nature} \textbf{\bibinfo{volume}{527}},
  \bibinfo{pages}{212} (\bibinfo{year}{2015}).

\bibitem[{\citenamefont{Suzuki et~al.}(2017)\citenamefont{Suzuki, Koretsune,
  Ochi, and Arita}}]{Suzuki_PhysRevB.95.094406}
\bibinfo{author}{\bibfnamefont{M.-T.} \bibnamefont{Suzuki}},
  \bibinfo{author}{\bibfnamefont{T.}~\bibnamefont{Koretsune}},
  \bibinfo{author}{\bibfnamefont{M.}~\bibnamefont{Ochi}}, \bibnamefont{and}
  \bibinfo{author}{\bibfnamefont{R.}~\bibnamefont{Arita}},
  \bibinfo{journal}{Phys. Rev. B} \textbf{\bibinfo{volume}{95}},
  \bibinfo{pages}{094406} (\bibinfo{year}{2017}).

\bibitem[{\citenamefont{Chen et~al.}(2020)\citenamefont{Chen, Wang, Xiao, Guo,
  Niu, and MacDonald}}]{Chen_PhysRevB.101.104418}
\bibinfo{author}{\bibfnamefont{H.}~\bibnamefont{Chen}},
  \bibinfo{author}{\bibfnamefont{T.-C.} \bibnamefont{Wang}},
  \bibinfo{author}{\bibfnamefont{D.}~\bibnamefont{Xiao}},
  \bibinfo{author}{\bibfnamefont{G.-Y.} \bibnamefont{Guo}},
  \bibinfo{author}{\bibfnamefont{Q.}~\bibnamefont{Niu}}, \bibnamefont{and}
  \bibinfo{author}{\bibfnamefont{A.~H.} \bibnamefont{MacDonald}},
  \bibinfo{journal}{Phys. Rev. B} \textbf{\bibinfo{volume}{101}},
  \bibinfo{pages}{104418} (\bibinfo{year}{2020}).

\bibitem[{\citenamefont{Ohgushi et~al.}(2000)\citenamefont{Ohgushi, Murakami,
  and Nagaosa}}]{Ohgushi_PhysRevB.62.R6065}
\bibinfo{author}{\bibfnamefont{K.}~\bibnamefont{Ohgushi}},
  \bibinfo{author}{\bibfnamefont{S.}~\bibnamefont{Murakami}}, \bibnamefont{and}
  \bibinfo{author}{\bibfnamefont{N.}~\bibnamefont{Nagaosa}},
  \bibinfo{journal}{Phys. Rev. B} \textbf{\bibinfo{volume}{62}},
  \bibinfo{pages}{R6065} (\bibinfo{year}{2000}).

\bibitem[{\citenamefont{Shindou and
  Nagaosa}(2001)}]{Shindou_PhysRevLett.87.116801}
\bibinfo{author}{\bibfnamefont{R.}~\bibnamefont{Shindou}} \bibnamefont{and}
  \bibinfo{author}{\bibfnamefont{N.}~\bibnamefont{Nagaosa}},
  \bibinfo{journal}{Phys. Rev. Lett.} \textbf{\bibinfo{volume}{87}},
  \bibinfo{pages}{116801} (\bibinfo{year}{2001}).

\bibitem[{\citenamefont{Nagaosa et~al.}(2010)\citenamefont{Nagaosa, Sinova,
  Onoda, MacDonald, and Ong}}]{Nagaosa_RevModPhys.82.1539}
\bibinfo{author}{\bibfnamefont{N.}~\bibnamefont{Nagaosa}},
  \bibinfo{author}{\bibfnamefont{J.}~\bibnamefont{Sinova}},
  \bibinfo{author}{\bibfnamefont{S.}~\bibnamefont{Onoda}},
  \bibinfo{author}{\bibfnamefont{A.~H.} \bibnamefont{MacDonald}},
  \bibnamefont{and} \bibinfo{author}{\bibfnamefont{N.~P.} \bibnamefont{Ong}},
  \bibinfo{journal}{Rev. Mod. Phys.} \textbf{\bibinfo{volume}{82}},
  \bibinfo{pages}{1539} (\bibinfo{year}{2010}).

\bibitem[{\citenamefont{Hlinka}(2014)}]{Hlinka_PhysRevLett.113.165502}
\bibinfo{author}{\bibfnamefont{J.}~\bibnamefont{Hlinka}},
  \bibinfo{journal}{Phys. Rev. Lett.} \textbf{\bibinfo{volume}{113}},
  \bibinfo{pages}{165502} (\bibinfo{year}{2014}).

\bibitem[{\citenamefont{Hlinka et~al.}(2016)\citenamefont{Hlinka, Privratska,
  Ondrejkovic, and Janovec}}]{Hlinka_PhysRevLett.116.177602}
\bibinfo{author}{\bibfnamefont{J.}~\bibnamefont{Hlinka}},
  \bibinfo{author}{\bibfnamefont{J.}~\bibnamefont{Privratska}},
  \bibinfo{author}{\bibfnamefont{P.}~\bibnamefont{Ondrejkovic}},
  \bibnamefont{and} \bibinfo{author}{\bibfnamefont{V.}~\bibnamefont{Janovec}},
  \bibinfo{journal}{Phys. Rev. Lett.} \textbf{\bibinfo{volume}{116}},
  \bibinfo{pages}{177602} (\bibinfo{year}{2016}).

\bibitem[{\citenamefont{Jin et~al.}(2020)\citenamefont{Jin, Drueke, Li, Admasu,
  Owen, Day, Sun, Cheong, and Zhao}}]{jin2020observation}
\bibinfo{author}{\bibfnamefont{W.}~\bibnamefont{Jin}},
  \bibinfo{author}{\bibfnamefont{E.}~\bibnamefont{Drueke}},
  \bibinfo{author}{\bibfnamefont{S.}~\bibnamefont{Li}},
  \bibinfo{author}{\bibfnamefont{A.}~\bibnamefont{Admasu}},
  \bibinfo{author}{\bibfnamefont{R.}~\bibnamefont{Owen}},
  \bibinfo{author}{\bibfnamefont{M.}~\bibnamefont{Day}},
  \bibinfo{author}{\bibfnamefont{K.}~\bibnamefont{Sun}},
  \bibinfo{author}{\bibfnamefont{S.-W.} \bibnamefont{Cheong}},
  \bibnamefont{and} \bibinfo{author}{\bibfnamefont{L.}~\bibnamefont{Zhao}},
  \bibinfo{journal}{Nat. Phys.} \textbf{\bibinfo{volume}{16}},
  \bibinfo{pages}{42} (\bibinfo{year}{2020}).

\bibitem[{\citenamefont{Hayashida et~al.}(2021)\citenamefont{Hayashida, Uemura,
  Kimura, Matsuoka, Hagihala, Hirose, Morioka, Hasegawa, and
  Kimura}}]{Hayashida_PhysRevMaterials.5.124409}
\bibinfo{author}{\bibfnamefont{T.}~\bibnamefont{Hayashida}},
  \bibinfo{author}{\bibfnamefont{Y.}~\bibnamefont{Uemura}},
  \bibinfo{author}{\bibfnamefont{K.}~\bibnamefont{Kimura}},
  \bibinfo{author}{\bibfnamefont{S.}~\bibnamefont{Matsuoka}},
  \bibinfo{author}{\bibfnamefont{M.}~\bibnamefont{Hagihala}},
  \bibinfo{author}{\bibfnamefont{S.}~\bibnamefont{Hirose}},
  \bibinfo{author}{\bibfnamefont{H.}~\bibnamefont{Morioka}},
  \bibinfo{author}{\bibfnamefont{T.}~\bibnamefont{Hasegawa}}, \bibnamefont{and}
  \bibinfo{author}{\bibfnamefont{T.}~\bibnamefont{Kimura}},
  \bibinfo{journal}{Phys. Rev. Materials} \textbf{\bibinfo{volume}{5}},
  \bibinfo{pages}{124409} (\bibinfo{year}{2021}).

\bibitem[{\citenamefont{Hayashida et~al.}(2020)\citenamefont{Hayashida, Uemura,
  Kimura, Matsuoka, Morikawa, Hirose, Tsuda, Hasegawa, and
  Kimura}}]{hayashida2020visualization}
\bibinfo{author}{\bibfnamefont{T.}~\bibnamefont{Hayashida}},
  \bibinfo{author}{\bibfnamefont{Y.}~\bibnamefont{Uemura}},
  \bibinfo{author}{\bibfnamefont{K.}~\bibnamefont{Kimura}},
  \bibinfo{author}{\bibfnamefont{S.}~\bibnamefont{Matsuoka}},
  \bibinfo{author}{\bibfnamefont{D.}~\bibnamefont{Morikawa}},
  \bibinfo{author}{\bibfnamefont{S.}~\bibnamefont{Hirose}},
  \bibinfo{author}{\bibfnamefont{K.}~\bibnamefont{Tsuda}},
  \bibinfo{author}{\bibfnamefont{T.}~\bibnamefont{Hasegawa}}, \bibnamefont{and}
  \bibinfo{author}{\bibfnamefont{T.}~\bibnamefont{Kimura}},
  \bibinfo{journal}{Nat. Commun.} \textbf{\bibinfo{volume}{11}},
  \bibinfo{pages}{4582} (\bibinfo{year}{2020}).

\bibitem[{\citenamefont{Yokota et~al.}(2022)\citenamefont{Yokota, Hayashida,
  Kitahara, and Kimura}}]{yokota2022three}
\bibinfo{author}{\bibfnamefont{H.}~\bibnamefont{Yokota}},
  \bibinfo{author}{\bibfnamefont{T.}~\bibnamefont{Hayashida}},
  \bibinfo{author}{\bibfnamefont{D.}~\bibnamefont{Kitahara}}, \bibnamefont{and}
  \bibinfo{author}{\bibfnamefont{T.}~\bibnamefont{Kimura}},
  \bibinfo{journal}{npj Quantum Mater.} \textbf{\bibinfo{volume}{7}},
  \bibinfo{pages}{106} (\bibinfo{year}{2022}).

\bibitem[{\citenamefont{Cheong et~al.}(2021)\citenamefont{Cheong, Lim, Du, and
  Huang}}]{cheong2021permutable}
\bibinfo{author}{\bibfnamefont{S.-W.} \bibnamefont{Cheong}},
  \bibinfo{author}{\bibfnamefont{S.}~\bibnamefont{Lim}},
  \bibinfo{author}{\bibfnamefont{K.}~\bibnamefont{Du}}, \bibnamefont{and}
  \bibinfo{author}{\bibfnamefont{F.-T.} \bibnamefont{Huang}},
  \bibinfo{journal}{npj Quantum Mater.} \textbf{\bibinfo{volume}{6}},
  \bibinfo{pages}{58} (\bibinfo{year}{2021}).

\bibitem[{\citenamefont{Hayami}(2022{\natexlab{a}})}]{Hayami_PhysRevB.106.144402}
\bibinfo{author}{\bibfnamefont{S.}~\bibnamefont{Hayami}},
  \bibinfo{journal}{Phys. Rev. B} \textbf{\bibinfo{volume}{106}},
  \bibinfo{pages}{144402} (\bibinfo{year}{2022}{\natexlab{a}}).

\bibitem[{\citenamefont{Cheong et~al.}(2022)\citenamefont{Cheong, Huang, and
  Kim}}]{cheong2022linking}
\bibinfo{author}{\bibfnamefont{S.-W.} \bibnamefont{Cheong}},
  \bibinfo{author}{\bibfnamefont{F.-T.} \bibnamefont{Huang}}, \bibnamefont{and}
  \bibinfo{author}{\bibfnamefont{M.}~\bibnamefont{Kim}}, \bibinfo{journal}{Rep.
  Prog. Phys.} \textbf{\bibinfo{volume}{85}}, \bibinfo{pages}{124501}
  (\bibinfo{year}{2022}).

\bibitem[{\citenamefont{Hayami et~al.}(2023{\natexlab{a}})\citenamefont{Hayami,
  Oiwa, and Kusunose}}]{hayami2023planar}
\bibinfo{author}{\bibfnamefont{S.}~\bibnamefont{Hayami}},
  \bibinfo{author}{\bibfnamefont{R.}~\bibnamefont{Oiwa}}, \bibnamefont{and}
  \bibinfo{author}{\bibfnamefont{H.}~\bibnamefont{Kusunose}},
  \bibinfo{journal}{arXiv:2306.12614}  (\bibinfo{year}{2023}{\natexlab{a}}).

\bibitem[{\citenamefont{Nasu and Hayami}(2022)}]{Nasu_PhysRevB.105.245125}
\bibinfo{author}{\bibfnamefont{J.}~\bibnamefont{Nasu}} \bibnamefont{and}
  \bibinfo{author}{\bibfnamefont{S.}~\bibnamefont{Hayami}},
  \bibinfo{journal}{Phys. Rev. B} \textbf{\bibinfo{volume}{105}},
  \bibinfo{pages}{245125} (\bibinfo{year}{2022}).

\bibitem[{\citenamefont{Roy et~al.}(2022)\citenamefont{Roy, Guimar\~aes, and
  S\l{}awi\ifmmode~\acute{n}\else
  \'{n}\fi{}ska}}]{Roy_PhysRevMaterials.6.045004}
\bibinfo{author}{\bibfnamefont{A.}~\bibnamefont{Roy}},
  \bibinfo{author}{\bibfnamefont{M.~H.~D.} \bibnamefont{Guimar\~aes}},
  \bibnamefont{and}
  \bibinfo{author}{\bibfnamefont{J.}~\bibnamefont{S\l{}awi\ifmmode~\acute{n}\else
  \'{n}\fi{}ska}}, \bibinfo{journal}{Phys. Rev. Materials}
  \textbf{\bibinfo{volume}{6}}, \bibinfo{pages}{045004} (\bibinfo{year}{2022}).

\bibitem[{\citenamefont{Hayami et~al.}(2022{\natexlab{a}})\citenamefont{Hayami,
  Oiwa, and Kusunose}}]{Hayami_doi:10.7566/JPSJ.91.113702}
\bibinfo{author}{\bibfnamefont{S.}~\bibnamefont{Hayami}},
  \bibinfo{author}{\bibfnamefont{R.}~\bibnamefont{Oiwa}}, \bibnamefont{and}
  \bibinfo{author}{\bibfnamefont{H.}~\bibnamefont{Kusunose}},
  \bibinfo{journal}{J. Phys. Soc. Jpn.} \textbf{\bibinfo{volume}{91}},
  \bibinfo{pages}{113702} (\bibinfo{year}{2022}{\natexlab{a}}).

\bibitem[{\citenamefont{Inda and Hayami}(2023)}]{inda2023nonlinear}
\bibinfo{author}{\bibfnamefont{A.}~\bibnamefont{Inda}} \bibnamefont{and}
  \bibinfo{author}{\bibfnamefont{S.}~\bibnamefont{Hayami}},
  \bibinfo{journal}{J. Phys. Soc. Jpn.} \textbf{\bibinfo{volume}{92}},
  \bibinfo{pages}{043701} (\bibinfo{year}{2023}).

\bibitem[{\citenamefont{Johnson et~al.}(2011)\citenamefont{Johnson, Nair,
  Chapon, Bombardi, Vecchini, Prabhakaran, Boothroyd, and
  Radaelli}}]{Johnson_PhysRevLett.107.137205}
\bibinfo{author}{\bibfnamefont{R.~D.} \bibnamefont{Johnson}},
  \bibinfo{author}{\bibfnamefont{S.}~\bibnamefont{Nair}},
  \bibinfo{author}{\bibfnamefont{L.~C.} \bibnamefont{Chapon}},
  \bibinfo{author}{\bibfnamefont{A.}~\bibnamefont{Bombardi}},
  \bibinfo{author}{\bibfnamefont{C.}~\bibnamefont{Vecchini}},
  \bibinfo{author}{\bibfnamefont{D.}~\bibnamefont{Prabhakaran}},
  \bibinfo{author}{\bibfnamefont{A.~T.} \bibnamefont{Boothroyd}},
  \bibnamefont{and} \bibinfo{author}{\bibfnamefont{P.~G.}
  \bibnamefont{Radaelli}}, \bibinfo{journal}{Phys. Rev. Lett.}
  \textbf{\bibinfo{volume}{107}}, \bibinfo{pages}{137205}
  (\bibinfo{year}{2011}).

\bibitem[{\citenamefont{Johnson et~al.}(2012)\citenamefont{Johnson, Chapon,
  Khalyavin, Manuel, Radaelli, and Martin}}]{Johnson_PhysRevLett.108.067201}
\bibinfo{author}{\bibfnamefont{R.~D.} \bibnamefont{Johnson}},
  \bibinfo{author}{\bibfnamefont{L.~C.} \bibnamefont{Chapon}},
  \bibinfo{author}{\bibfnamefont{D.~D.} \bibnamefont{Khalyavin}},
  \bibinfo{author}{\bibfnamefont{P.}~\bibnamefont{Manuel}},
  \bibinfo{author}{\bibfnamefont{P.~G.} \bibnamefont{Radaelli}},
  \bibnamefont{and} \bibinfo{author}{\bibfnamefont{C.}~\bibnamefont{Martin}},
  \bibinfo{journal}{Phys. Rev. Lett.} \textbf{\bibinfo{volume}{108}},
  \bibinfo{pages}{067201} (\bibinfo{year}{2012}).

\bibitem[{\citenamefont{Hasegawa et~al.}(2020)\citenamefont{Hasegawa, Yoshida,
  Nakamura, Ogita, and Matsuhira}}]{Hasegawa_doi:10.7566/JPSJ.89.054602}
\bibinfo{author}{\bibfnamefont{T.}~\bibnamefont{Hasegawa}},
  \bibinfo{author}{\bibfnamefont{W.}~\bibnamefont{Yoshida}},
  \bibinfo{author}{\bibfnamefont{K.}~\bibnamefont{Nakamura}},
  \bibinfo{author}{\bibfnamefont{N.}~\bibnamefont{Ogita}}, \bibnamefont{and}
  \bibinfo{author}{\bibfnamefont{K.}~\bibnamefont{Matsuhira}},
  \bibinfo{journal}{J. Phys. Soc. Jpn.} \textbf{\bibinfo{volume}{89}},
  \bibinfo{pages}{054602} (\bibinfo{year}{2020}).

\bibitem[{\citenamefont{Hanate et~al.}(2021)\citenamefont{Hanate, Hasegawa,
  Hayami, Tsutsui, Kawano, and Matsuhira}}]{hanate2021first}
\bibinfo{author}{\bibfnamefont{H.}~\bibnamefont{Hanate}},
  \bibinfo{author}{\bibfnamefont{T.}~\bibnamefont{Hasegawa}},
  \bibinfo{author}{\bibfnamefont{S.}~\bibnamefont{Hayami}},
  \bibinfo{author}{\bibfnamefont{S.}~\bibnamefont{Tsutsui}},
  \bibinfo{author}{\bibfnamefont{S.}~\bibnamefont{Kawano}}, \bibnamefont{and}
  \bibinfo{author}{\bibfnamefont{K.}~\bibnamefont{Matsuhira}},
  \bibinfo{journal}{J. Phys. Soc. Jpn.} \textbf{\bibinfo{volume}{90}},
  \bibinfo{pages}{063702} (\bibinfo{year}{2021}).

\bibitem[{\citenamefont{Hayami et~al.}(2023{\natexlab{b}})\citenamefont{Hayami,
  Tsutsui, Hanate, Nagasawa, Yoda, and Matsuhira}}]{hayami2023cluster}
\bibinfo{author}{\bibfnamefont{S.}~\bibnamefont{Hayami}},
  \bibinfo{author}{\bibfnamefont{S.}~\bibnamefont{Tsutsui}},
  \bibinfo{author}{\bibfnamefont{H.}~\bibnamefont{Hanate}},
  \bibinfo{author}{\bibfnamefont{N.}~\bibnamefont{Nagasawa}},
  \bibinfo{author}{\bibfnamefont{Y.}~\bibnamefont{Yoda}}, \bibnamefont{and}
  \bibinfo{author}{\bibfnamefont{K.}~\bibnamefont{Matsuhira}},
  \bibinfo{journal}{J. Phys. Soc. Jpn.} \textbf{\bibinfo{volume}{92}},
  \bibinfo{pages}{033702} (\bibinfo{year}{2023}{\natexlab{b}}).

\bibitem[{\citenamefont{Hanate et~al.}(2023)\citenamefont{Hanate, Tsutsui,
  Yajima, Nakao, Sagayama, Hasegawa, and Matsuhira}}]{hanate2023space}
\bibinfo{author}{\bibfnamefont{H.}~\bibnamefont{Hanate}},
  \bibinfo{author}{\bibfnamefont{S.}~\bibnamefont{Tsutsui}},
  \bibinfo{author}{\bibfnamefont{T.}~\bibnamefont{Yajima}},
  \bibinfo{author}{\bibfnamefont{H.}~\bibnamefont{Nakao}},
  \bibinfo{author}{\bibfnamefont{H.}~\bibnamefont{Sagayama}},
  \bibinfo{author}{\bibfnamefont{T.}~\bibnamefont{Hasegawa}}, \bibnamefont{and}
  \bibinfo{author}{\bibfnamefont{K.}~\bibnamefont{Matsuhira}},
  \bibinfo{journal}{J. Phys. Soc. Jpn.} \textbf{\bibinfo{volume}{92}},
  \bibinfo{pages}{063601} (\bibinfo{year}{2023}).

\bibitem[{\citenamefont{Xu et~al.}(2022)\citenamefont{Xu, Huang, Admasu,
  Kratochv\'{\i}lov\'a, Chu, Park, and Cheong}}]{Xu_PhysRevB.105.184407}
\bibinfo{author}{\bibfnamefont{X.}~\bibnamefont{Xu}},
  \bibinfo{author}{\bibfnamefont{F.-T.} \bibnamefont{Huang}},
  \bibinfo{author}{\bibfnamefont{A.~S.} \bibnamefont{Admasu}},
  \bibinfo{author}{\bibfnamefont{M.}~\bibnamefont{Kratochv\'{\i}lov\'a}},
  \bibinfo{author}{\bibfnamefont{M.-W.} \bibnamefont{Chu}},
  \bibinfo{author}{\bibfnamefont{J.-G.} \bibnamefont{Park}}, \bibnamefont{and}
  \bibinfo{author}{\bibfnamefont{S.-W.} \bibnamefont{Cheong}},
  \bibinfo{journal}{Phys. Rev. B} \textbf{\bibinfo{volume}{105}},
  \bibinfo{pages}{184407} (\bibinfo{year}{2022}).

\bibitem[{\citenamefont{Yamagishi et~al.}(2023)\citenamefont{Yamagishi,
  Hayashida, Misawa, Kimura, Hagihala, Murata, Hirose, and
  Kimura}}]{yamagishi2023ferroaxial}
\bibinfo{author}{\bibfnamefont{S.}~\bibnamefont{Yamagishi}},
  \bibinfo{author}{\bibfnamefont{T.}~\bibnamefont{Hayashida}},
  \bibinfo{author}{\bibfnamefont{R.}~\bibnamefont{Misawa}},
  \bibinfo{author}{\bibfnamefont{K.}~\bibnamefont{Kimura}},
  \bibinfo{author}{\bibfnamefont{M.}~\bibnamefont{Hagihala}},
  \bibinfo{author}{\bibfnamefont{T.}~\bibnamefont{Murata}},
  \bibinfo{author}{\bibfnamefont{S.}~\bibnamefont{Hirose}}, \bibnamefont{and}
  \bibinfo{author}{\bibfnamefont{T.}~\bibnamefont{Kimura}},
  \bibinfo{journal}{Chem. Mater.} \textbf{\bibinfo{volume}{35}},
  \bibinfo{pages}{747} (\bibinfo{year}{2023}).

\bibitem[{\citenamefont{Nagai and Kimura}(2023)}]{nagai2023chemicalSwitching}
\bibinfo{author}{\bibfnamefont{T.}~\bibnamefont{Nagai}} \bibnamefont{and}
  \bibinfo{author}{\bibfnamefont{T.}~\bibnamefont{Kimura}},
  \bibinfo{journal}{Chem. Mater.}  (\bibinfo{year}{2023}).

\bibitem[{\citenamefont{Nagai et~al.}(2023)\citenamefont{Nagai, Mochizuki,
  Yoshida, and Kimura}}]{nagai2023chemical}
\bibinfo{author}{\bibfnamefont{T.}~\bibnamefont{Nagai}},
  \bibinfo{author}{\bibfnamefont{Y.}~\bibnamefont{Mochizuki}},
  \bibinfo{author}{\bibfnamefont{S.}~\bibnamefont{Yoshida}}, \bibnamefont{and}
  \bibinfo{author}{\bibfnamefont{T.}~\bibnamefont{Kimura}},
  \bibinfo{journal}{J. Am. Chem. Soc.} \textbf{\bibinfo{volume}{145}},
  \bibinfo{pages}{8090} (\bibinfo{year}{2023}).

\bibitem[{\citenamefont{Yanase}(2014)}]{Yanase_JPSJ.83.014703}
\bibinfo{author}{\bibfnamefont{Y.}~\bibnamefont{Yanase}}, \bibinfo{journal}{J.
  Phys. Soc. Jpn.} \textbf{\bibinfo{volume}{83}}, \bibinfo{pages}{014703}
  (\bibinfo{year}{2014}).

\bibitem[{\citenamefont{Hayami et~al.}(2015)\citenamefont{Hayami, Kusunose, and
  Motome}}]{Hayami_doi:10.7566/JPSJ.84.064717}
\bibinfo{author}{\bibfnamefont{S.}~\bibnamefont{Hayami}},
  \bibinfo{author}{\bibfnamefont{H.}~\bibnamefont{Kusunose}}, \bibnamefont{and}
  \bibinfo{author}{\bibfnamefont{Y.}~\bibnamefont{Motome}},
  \bibinfo{journal}{J. Phys. Soc. Jpn.} \textbf{\bibinfo{volume}{84}},
  \bibinfo{pages}{064717} (\bibinfo{year}{2015}).

\bibitem[{\citenamefont{Kusunose and
  Hayami}(2022)}]{kusunose2022generalization}
\bibinfo{author}{\bibfnamefont{H.}~\bibnamefont{Kusunose}} \bibnamefont{and}
  \bibinfo{author}{\bibfnamefont{S.}~\bibnamefont{Hayami}},
  \bibinfo{journal}{J. Phys.: Condens. Matter} \textbf{\bibinfo{volume}{34}},
  \bibinfo{pages}{464002} (\bibinfo{year}{2022}).

\bibitem[{\citenamefont{Hayami et~al.}(2014)\citenamefont{Hayami, Kusunose, and
  Motome}}]{Hayami_PhysRevB.90.081115}
\bibinfo{author}{\bibfnamefont{S.}~\bibnamefont{Hayami}},
  \bibinfo{author}{\bibfnamefont{H.}~\bibnamefont{Kusunose}}, \bibnamefont{and}
  \bibinfo{author}{\bibfnamefont{Y.}~\bibnamefont{Motome}},
  \bibinfo{journal}{Phys. Rev. B} \textbf{\bibinfo{volume}{90}},
  \bibinfo{pages}{081115} (\bibinfo{year}{2014}).

\bibitem[{\citenamefont{Hayami et~al.}(2016{\natexlab{a}})\citenamefont{Hayami,
  Kusunose, and Motome}}]{hayami2016emergent}
\bibinfo{author}{\bibfnamefont{S.}~\bibnamefont{Hayami}},
  \bibinfo{author}{\bibfnamefont{H.}~\bibnamefont{Kusunose}}, \bibnamefont{and}
  \bibinfo{author}{\bibfnamefont{Y.}~\bibnamefont{Motome}},
  \bibinfo{journal}{J. Phys.: Condens. Matter} \textbf{\bibinfo{volume}{28}},
  \bibinfo{pages}{395601} (\bibinfo{year}{2016}{\natexlab{a}}).

\bibitem[{\citenamefont{Watanabe and
  Yanase}(2017)}]{Watanabe_PhysRevB.96.064432}
\bibinfo{author}{\bibfnamefont{H.}~\bibnamefont{Watanabe}} \bibnamefont{and}
  \bibinfo{author}{\bibfnamefont{Y.}~\bibnamefont{Yanase}},
  \bibinfo{journal}{Phys. Rev. B} \textbf{\bibinfo{volume}{96}},
  \bibinfo{pages}{064432} (\bibinfo{year}{2017}).

\bibitem[{\citenamefont{Hayami and
  Kusunose}(2021{\natexlab{b}})}]{Hayami_PhysRevB.104.045117}
\bibinfo{author}{\bibfnamefont{S.}~\bibnamefont{Hayami}} \bibnamefont{and}
  \bibinfo{author}{\bibfnamefont{H.}~\bibnamefont{Kusunose}},
  \bibinfo{journal}{Phys. Rev. B} \textbf{\bibinfo{volume}{104}},
  \bibinfo{pages}{045117} (\bibinfo{year}{2021}{\natexlab{b}}).

\bibitem[{\citenamefont{Kirikoshi and
  Hayami}(2023)}]{Kirikoshi_PhysRevB.107.155109}
\bibinfo{author}{\bibfnamefont{A.}~\bibnamefont{Kirikoshi}} \bibnamefont{and}
  \bibinfo{author}{\bibfnamefont{S.}~\bibnamefont{Hayami}},
  \bibinfo{journal}{Phys. Rev. B} \textbf{\bibinfo{volume}{107}},
  \bibinfo{pages}{155109} (\bibinfo{year}{2023}).

\bibitem[{\citenamefont{Hayami}(2022{\natexlab{b}})}]{Hayami_PhysRevB.105.014408}
\bibinfo{author}{\bibfnamefont{S.}~\bibnamefont{Hayami}},
  \bibinfo{journal}{Phys. Rev. B} \textbf{\bibinfo{volume}{105}},
  \bibinfo{pages}{014408} (\bibinfo{year}{2022}{\natexlab{b}}).

\bibitem[{\citenamefont{Hayami}(2022{\natexlab{c}})}]{Hayami_PhysRevB.105.184426}
\bibinfo{author}{\bibfnamefont{S.}~\bibnamefont{Hayami}},
  \bibinfo{journal}{Phys. Rev. B} \textbf{\bibinfo{volume}{105}},
  \bibinfo{pages}{184426} (\bibinfo{year}{2022}{\natexlab{c}}).

\bibitem[{\citenamefont{Hayami}(2022{\natexlab{d}})}]{hayami2022square}
\bibinfo{author}{\bibfnamefont{S.}~\bibnamefont{Hayami}}, \bibinfo{journal}{J.
  Phys.: Condens. Matter} \textbf{\bibinfo{volume}{34}},
  \bibinfo{pages}{365802} (\bibinfo{year}{2022}{\natexlab{d}}).

\bibitem[{\citenamefont{Lin}(2021)}]{lin2021skyrmion}
\bibinfo{author}{\bibfnamefont{S.-Z.} \bibnamefont{Lin}},
  \bibinfo{journal}{arXiv:2112.12850}  (\bibinfo{year}{2021}).

\bibitem[{\citenamefont{Maruyama et~al.}(2012)\citenamefont{Maruyama, Sigrist,
  and Yanase}}]{Maruyama_doi:10.1143/JPSJ.81.034702}
\bibinfo{author}{\bibfnamefont{D.}~\bibnamefont{Maruyama}},
  \bibinfo{author}{\bibfnamefont{M.}~\bibnamefont{Sigrist}}, \bibnamefont{and}
  \bibinfo{author}{\bibfnamefont{Y.}~\bibnamefont{Yanase}},
  \bibinfo{journal}{J. Phys. Soc. Jpn.} \textbf{\bibinfo{volume}{81}},
  \bibinfo{pages}{034702} (\bibinfo{year}{2012}).

\bibitem[{\citenamefont{Goryo et~al.}(2012)\citenamefont{Goryo, Fischer, and
  Sigrist}}]{Goryo_PhysRevB.86.100507}
\bibinfo{author}{\bibfnamefont{J.}~\bibnamefont{Goryo}},
  \bibinfo{author}{\bibfnamefont{M.~H.} \bibnamefont{Fischer}},
  \bibnamefont{and} \bibinfo{author}{\bibfnamefont{M.}~\bibnamefont{Sigrist}},
  \bibinfo{journal}{Phys. Rev. B} \textbf{\bibinfo{volume}{86}},
  \bibinfo{pages}{100507} (\bibinfo{year}{2012}).

\bibitem[{\citenamefont{Yoshida et~al.}(2012)\citenamefont{Yoshida, Sigrist,
  and Yanase}}]{Yoshida_PhysRevB.86.134514}
\bibinfo{author}{\bibfnamefont{T.}~\bibnamefont{Yoshida}},
  \bibinfo{author}{\bibfnamefont{M.}~\bibnamefont{Sigrist}}, \bibnamefont{and}
  \bibinfo{author}{\bibfnamefont{Y.}~\bibnamefont{Yanase}},
  \bibinfo{journal}{Phys. Rev. B} \textbf{\bibinfo{volume}{86}},
  \bibinfo{pages}{134514} (\bibinfo{year}{2012}).

\bibitem[{\citenamefont{Yoshida
  et~al.}(2013{\natexlab{a}})\citenamefont{Yoshida, Sigrist, and
  Yanase}}]{yoshida2013parity}
\bibinfo{author}{\bibfnamefont{T.}~\bibnamefont{Yoshida}},
  \bibinfo{author}{\bibfnamefont{M.}~\bibnamefont{Sigrist}}, \bibnamefont{and}
  \bibinfo{author}{\bibfnamefont{Y.}~\bibnamefont{Yanase}},
  \bibinfo{journal}{J. Phys. Soc. Jpn.} \textbf{\bibinfo{volume}{83}},
  \bibinfo{pages}{013703} (\bibinfo{year}{2013}{\natexlab{a}}).

\bibitem[{\citenamefont{Yoshida
  et~al.}(2013{\natexlab{b}})\citenamefont{Yoshida, Sigrist, and
  Yanase}}]{Yoshida_doi:10.7566/JPSJ.82.074714}
\bibinfo{author}{\bibfnamefont{T.}~\bibnamefont{Yoshida}},
  \bibinfo{author}{\bibfnamefont{M.}~\bibnamefont{Sigrist}}, \bibnamefont{and}
  \bibinfo{author}{\bibfnamefont{Y.}~\bibnamefont{Yanase}},
  \bibinfo{journal}{J. Phys. Soc. Jpn.} \textbf{\bibinfo{volume}{82}},
  \bibinfo{pages}{074714} (\bibinfo{year}{2013}{\natexlab{b}}).

\bibitem[{\citenamefont{Sigrist et~al.}(2014)\citenamefont{Sigrist, Agterberg,
  Fischer, Goryo, Loder, Rhim, Maruyama, Yanase, Yoshida, and
  Youn}}]{sigrist2014superconductors}
\bibinfo{author}{\bibfnamefont{M.}~\bibnamefont{Sigrist}},
  \bibinfo{author}{\bibfnamefont{D.~F.} \bibnamefont{Agterberg}},
  \bibinfo{author}{\bibfnamefont{M.~H.} \bibnamefont{Fischer}},
  \bibinfo{author}{\bibfnamefont{J.}~\bibnamefont{Goryo}},
  \bibinfo{author}{\bibfnamefont{F.}~\bibnamefont{Loder}},
  \bibinfo{author}{\bibfnamefont{S.-H.} \bibnamefont{Rhim}},
  \bibinfo{author}{\bibfnamefont{D.}~\bibnamefont{Maruyama}},
  \bibinfo{author}{\bibfnamefont{Y.}~\bibnamefont{Yanase}},
  \bibinfo{author}{\bibfnamefont{T.}~\bibnamefont{Yoshida}}, \bibnamefont{and}
  \bibinfo{author}{\bibfnamefont{S.~J.} \bibnamefont{Youn}},
  \bibinfo{journal}{J. Phys. Soc. Jpn.} \textbf{\bibinfo{volume}{83}},
  \bibinfo{pages}{061014} (\bibinfo{year}{2014}).

\bibitem[{\citenamefont{Yoshida et~al.}(2015)\citenamefont{Yoshida, Sigrist,
  and Yanase}}]{Yoshida_PhysRevLett.115.027001}
\bibinfo{author}{\bibfnamefont{T.}~\bibnamefont{Yoshida}},
  \bibinfo{author}{\bibfnamefont{M.}~\bibnamefont{Sigrist}}, \bibnamefont{and}
  \bibinfo{author}{\bibfnamefont{Y.}~\bibnamefont{Yanase}},
  \bibinfo{journal}{Phys. Rev. Lett.} \textbf{\bibinfo{volume}{115}},
  \bibinfo{pages}{027001} (\bibinfo{year}{2015}).

\bibitem[{\citenamefont{Sumita and Yanase}(2016)}]{Sumita_PhysRevB.93.224507}
\bibinfo{author}{\bibfnamefont{S.}~\bibnamefont{Sumita}} \bibnamefont{and}
  \bibinfo{author}{\bibfnamefont{Y.}~\bibnamefont{Yanase}},
  \bibinfo{journal}{Phys. Rev. B} \textbf{\bibinfo{volume}{93}},
  \bibinfo{pages}{224507} (\bibinfo{year}{2016}).

\bibitem[{\citenamefont{Ishizuka and
  Yanase}(2018)}]{Ishizuka_PhysRevB.98.224510}
\bibinfo{author}{\bibfnamefont{J.}~\bibnamefont{Ishizuka}} \bibnamefont{and}
  \bibinfo{author}{\bibfnamefont{Y.}~\bibnamefont{Yanase}},
  \bibinfo{journal}{Phys. Rev. B} \textbf{\bibinfo{volume}{98}},
  \bibinfo{pages}{224510} (\bibinfo{year}{2018}).

\bibitem[{\citenamefont{Nogaki et~al.}(2021)\citenamefont{Nogaki, Daido,
  Ishizuka, and Yanase}}]{Nogaki_PhysRevResearch.3.L032071}
\bibinfo{author}{\bibfnamefont{K.}~\bibnamefont{Nogaki}},
  \bibinfo{author}{\bibfnamefont{A.}~\bibnamefont{Daido}},
  \bibinfo{author}{\bibfnamefont{J.}~\bibnamefont{Ishizuka}}, \bibnamefont{and}
  \bibinfo{author}{\bibfnamefont{Y.}~\bibnamefont{Yanase}},
  \bibinfo{journal}{Phys. Rev. Res.} \textbf{\bibinfo{volume}{3}},
  \bibinfo{pages}{L032071} (\bibinfo{year}{2021}).

\bibitem[{\citenamefont{M\"ockli and
  Ramires}(2021)}]{Mockli_PhysRevB.104.134517}
\bibinfo{author}{\bibfnamefont{D.}~\bibnamefont{M\"ockli}} \bibnamefont{and}
  \bibinfo{author}{\bibfnamefont{A.}~\bibnamefont{Ramires}},
  \bibinfo{journal}{Phys. Rev. B} \textbf{\bibinfo{volume}{104}},
  \bibinfo{pages}{134517} (\bibinfo{year}{2021}).

\bibitem[{\citenamefont{Fischer et~al.}(2023)\citenamefont{Fischer, Sigrist,
  Agterberg, and Yanase}}]{fischer2023superconductivity}
\bibinfo{author}{\bibfnamefont{M.~H.} \bibnamefont{Fischer}},
  \bibinfo{author}{\bibfnamefont{M.}~\bibnamefont{Sigrist}},
  \bibinfo{author}{\bibfnamefont{D.~F.} \bibnamefont{Agterberg}},
  \bibnamefont{and} \bibinfo{author}{\bibfnamefont{Y.}~\bibnamefont{Yanase}},
  \bibinfo{journal}{Annu. Rev. Condens. Matter Phys.}
  \textbf{\bibinfo{volume}{14}}, \bibinfo{pages}{153} (\bibinfo{year}{2023}).

\bibitem[{\citenamefont{Cysne et~al.}(2021)\citenamefont{Cysne, Guimar\~aes,
  Canonico, Rappoport, and Muniz}}]{cysne2021orbital}
\bibinfo{author}{\bibfnamefont{T.~P.} \bibnamefont{Cysne}},
  \bibinfo{author}{\bibfnamefont{F.~S.~M.} \bibnamefont{Guimar\~aes}},
  \bibinfo{author}{\bibfnamefont{L.~M.} \bibnamefont{Canonico}},
  \bibinfo{author}{\bibfnamefont{T.~G.} \bibnamefont{Rappoport}},
  \bibnamefont{and} \bibinfo{author}{\bibfnamefont{R.~B.} \bibnamefont{Muniz}},
  \bibinfo{journal}{Phys. Rev. B} \textbf{\bibinfo{volume}{104}},
  \bibinfo{pages}{165403} (\bibinfo{year}{2021}).

\bibitem[{\citenamefont{Suzuki}(2022)}]{Suzuki_PhysRevB.105.075201}
\bibinfo{author}{\bibfnamefont{Y.}~\bibnamefont{Suzuki}},
  \bibinfo{journal}{Phys. Rev. B} \textbf{\bibinfo{volume}{105}},
  \bibinfo{pages}{075201} (\bibinfo{year}{2022}).

\bibitem[{\citenamefont{Yatsushiro et~al.}(2022)\citenamefont{Yatsushiro, Oiwa,
  Kusunose, and Hayami}}]{Yatsushiro_PhysRevB.105.155157}
\bibinfo{author}{\bibfnamefont{M.}~\bibnamefont{Yatsushiro}},
  \bibinfo{author}{\bibfnamefont{R.}~\bibnamefont{Oiwa}},
  \bibinfo{author}{\bibfnamefont{H.}~\bibnamefont{Kusunose}}, \bibnamefont{and}
  \bibinfo{author}{\bibfnamefont{S.}~\bibnamefont{Hayami}},
  \bibinfo{journal}{Phys. Rev. B} \textbf{\bibinfo{volume}{105}},
  \bibinfo{pages}{155157} (\bibinfo{year}{2022}).

\bibitem[{\citenamefont{Kondo and
  Akagi}(2022)}]{Kondo_PhysRevResearch.4.013186}
\bibinfo{author}{\bibfnamefont{H.}~\bibnamefont{Kondo}} \bibnamefont{and}
  \bibinfo{author}{\bibfnamefont{Y.}~\bibnamefont{Akagi}},
  \bibinfo{journal}{Phys. Rev. Research} \textbf{\bibinfo{volume}{4}},
  \bibinfo{pages}{013186} (\bibinfo{year}{2022}).

\bibitem[{\citenamefont{Hayami et~al.}(2022{\natexlab{b}})\citenamefont{Hayami,
  Yatsushiro, and Kusunose}}]{Hayami_PhysRevB.106.024405}
\bibinfo{author}{\bibfnamefont{S.}~\bibnamefont{Hayami}},
  \bibinfo{author}{\bibfnamefont{M.}~\bibnamefont{Yatsushiro}},
  \bibnamefont{and} \bibinfo{author}{\bibfnamefont{H.}~\bibnamefont{Kusunose}},
  \bibinfo{journal}{Phys. Rev. B} \textbf{\bibinfo{volume}{106}},
  \bibinfo{pages}{024405} (\bibinfo{year}{2022}{\natexlab{b}}).

\bibitem[{\citenamefont{Hayami et~al.}(2016{\natexlab{b}})\citenamefont{Hayami,
  Kusunose, and Motome}}]{Hayami_doi:10.7566/JPSJ.85.053705}
\bibinfo{author}{\bibfnamefont{S.}~\bibnamefont{Hayami}},
  \bibinfo{author}{\bibfnamefont{H.}~\bibnamefont{Kusunose}}, \bibnamefont{and}
  \bibinfo{author}{\bibfnamefont{Y.}~\bibnamefont{Motome}},
  \bibinfo{journal}{J. Phys. Soc. Jpn.} \textbf{\bibinfo{volume}{85}},
  \bibinfo{pages}{053705} (\bibinfo{year}{2016}{\natexlab{b}}).

\bibitem[{\citenamefont{Takashima et~al.}(2018)\citenamefont{Takashima, Shiomi,
  and Motome}}]{Takashima_PhysRevB.98.020401}
\bibinfo{author}{\bibfnamefont{R.}~\bibnamefont{Takashima}},
  \bibinfo{author}{\bibfnamefont{Y.}~\bibnamefont{Shiomi}}, \bibnamefont{and}
  \bibinfo{author}{\bibfnamefont{Y.}~\bibnamefont{Motome}},
  \bibinfo{journal}{Phys. Rev. B} \textbf{\bibinfo{volume}{98}},
  \bibinfo{pages}{020401(R)} (\bibinfo{year}{2018}).

\bibitem[{\citenamefont{Matsumoto and
  Hayami}(2020)}]{Matsumoto_PhysRevB.101.224419}
\bibinfo{author}{\bibfnamefont{T.}~\bibnamefont{Matsumoto}} \bibnamefont{and}
  \bibinfo{author}{\bibfnamefont{S.}~\bibnamefont{Hayami}},
  \bibinfo{journal}{Phys. Rev. B} \textbf{\bibinfo{volume}{101}},
  \bibinfo{pages}{224419} (\bibinfo{year}{2020}).

\bibitem[{\citenamefont{Matsumoto and
  Hayami}(2021)}]{Matsumoto_PhysRevB.104.134420}
\bibinfo{author}{\bibfnamefont{T.}~\bibnamefont{Matsumoto}} \bibnamefont{and}
  \bibinfo{author}{\bibfnamefont{S.}~\bibnamefont{Hayami}},
  \bibinfo{journal}{Phys. Rev. B} \textbf{\bibinfo{volume}{104}},
  \bibinfo{pages}{134420} (\bibinfo{year}{2021}).

\bibitem[{\citenamefont{Hayami and
  Matsumoto}(2022)}]{Hayami_PhysRevB.105.014404}
\bibinfo{author}{\bibfnamefont{S.}~\bibnamefont{Hayami}} \bibnamefont{and}
  \bibinfo{author}{\bibfnamefont{T.}~\bibnamefont{Matsumoto}},
  \bibinfo{journal}{Phys. Rev. B} \textbf{\bibinfo{volume}{105}},
  \bibinfo{pages}{014404} (\bibinfo{year}{2022}).

\bibitem[{\citenamefont{Kane and Mele}(2005)}]{Kane_PhysRevLett.95.226801}
\bibinfo{author}{\bibfnamefont{C.~L.} \bibnamefont{Kane}} \bibnamefont{and}
  \bibinfo{author}{\bibfnamefont{E.~J.} \bibnamefont{Mele}},
  \bibinfo{journal}{Phys. Rev. Lett.} \textbf{\bibinfo{volume}{95}},
  \bibinfo{pages}{226801} (\bibinfo{year}{2005}).

\bibitem[{\citenamefont{Yanagi and Kusunose}(2017)}]{yanagi2017optical}
\bibinfo{author}{\bibfnamefont{Y.}~\bibnamefont{Yanagi}} \bibnamefont{and}
  \bibinfo{author}{\bibfnamefont{H.}~\bibnamefont{Kusunose}},
  \bibinfo{journal}{J. Phys. Soc. Jpn.} \textbf{\bibinfo{volume}{86}},
  \bibinfo{pages}{083703} (\bibinfo{year}{2017}).

\bibitem[{\citenamefont{Yanagi et~al.}(2018)\citenamefont{Yanagi, Hayami, and
  Kusunose}}]{Yanagi_PhysRevB.97.020404}
\bibinfo{author}{\bibfnamefont{Y.}~\bibnamefont{Yanagi}},
  \bibinfo{author}{\bibfnamefont{S.}~\bibnamefont{Hayami}}, \bibnamefont{and}
  \bibinfo{author}{\bibfnamefont{H.}~\bibnamefont{Kusunose}},
  \bibinfo{journal}{Phys. Rev. B} \textbf{\bibinfo{volume}{97}},
  \bibinfo{pages}{020404} (\bibinfo{year}{2018}).

\bibitem[{\citenamefont{Fu et~al.}(2007)\citenamefont{Fu, Kane, and
  Mele}}]{Fu_PhysRevLett.98.106803}
\bibinfo{author}{\bibfnamefont{L.}~\bibnamefont{Fu}},
  \bibinfo{author}{\bibfnamefont{C.~L.} \bibnamefont{Kane}}, \bibnamefont{and}
  \bibinfo{author}{\bibfnamefont{E.~J.} \bibnamefont{Mele}},
  \bibinfo{journal}{Phys. Rev. Lett.} \textbf{\bibinfo{volume}{98}},
  \bibinfo{pages}{106803} (\bibinfo{year}{2007}).

\bibitem[{\citenamefont{Hayami et~al.}(2018{\natexlab{a}})\citenamefont{Hayami,
  Kusunose, and Motome}}]{Hayami_PhysRevB.97.024414}
\bibinfo{author}{\bibfnamefont{S.}~\bibnamefont{Hayami}},
  \bibinfo{author}{\bibfnamefont{H.}~\bibnamefont{Kusunose}}, \bibnamefont{and}
  \bibinfo{author}{\bibfnamefont{Y.}~\bibnamefont{Motome}},
  \bibinfo{journal}{Phys. Rev. B} \textbf{\bibinfo{volume}{97}},
  \bibinfo{pages}{024414} (\bibinfo{year}{2018}{\natexlab{a}}).

\bibitem[{\citenamefont{Ishitobi and
  Hattori}(2019)}]{Ishitobi_doi:10.7566/JPSJ.88.063708}
\bibinfo{author}{\bibfnamefont{T.}~\bibnamefont{Ishitobi}} \bibnamefont{and}
  \bibinfo{author}{\bibfnamefont{K.}~\bibnamefont{Hattori}},
  \bibinfo{journal}{J. Phys. Soc. Jpn.} \textbf{\bibinfo{volume}{88}},
  \bibinfo{pages}{063708} (\bibinfo{year}{2019}).

\bibitem[{\citenamefont{Hitomi and Yanase}(2014)}]{hitomi2014electric}
\bibinfo{author}{\bibfnamefont{T.}~\bibnamefont{Hitomi}} \bibnamefont{and}
  \bibinfo{author}{\bibfnamefont{Y.}~\bibnamefont{Yanase}},
  \bibinfo{journal}{J. Phys. Soc. Jpn.} \textbf{\bibinfo{volume}{83}},
  \bibinfo{pages}{114704} (\bibinfo{year}{2014}).

\bibitem[{\citenamefont{Hitomi and Yanase}(2016)}]{hitomi2016electric}
\bibinfo{author}{\bibfnamefont{T.}~\bibnamefont{Hitomi}} \bibnamefont{and}
  \bibinfo{author}{\bibfnamefont{Y.}~\bibnamefont{Yanase}},
  \bibinfo{journal}{J. Phys. Soc. Jpn.} \textbf{\bibinfo{volume}{85}},
  \bibinfo{pages}{124702} (\bibinfo{year}{2016}).

\bibitem[{\citenamefont{Yatsushiro and
  Hayami}(2020{\natexlab{a}})}]{yatsushiro2020odd}
\bibinfo{author}{\bibfnamefont{M.}~\bibnamefont{Yatsushiro}} \bibnamefont{and}
  \bibinfo{author}{\bibfnamefont{S.}~\bibnamefont{Hayami}},
  \bibinfo{journal}{J. Phys. Soc. Jpn.} \textbf{\bibinfo{volume}{89}},
  \bibinfo{pages}{013703} (\bibinfo{year}{2020}{\natexlab{a}}).

\bibitem[{\citenamefont{Yatsushiro and
  Hayami}(2020{\natexlab{b}})}]{Yatsushiro_PhysRevB.102.195147}
\bibinfo{author}{\bibfnamefont{M.}~\bibnamefont{Yatsushiro}} \bibnamefont{and}
  \bibinfo{author}{\bibfnamefont{S.}~\bibnamefont{Hayami}},
  \bibinfo{journal}{Phys. Rev. B} \textbf{\bibinfo{volume}{102}},
  \bibinfo{pages}{195147} (\bibinfo{year}{2020}{\natexlab{b}}).

\bibitem[{\citenamefont{Saito et~al.}(2018)\citenamefont{Saito, Uenishi, Miura,
  Tabata, Hidaka, Yanagisawa, and Amitsuka}}]{saito2018evidence}
\bibinfo{author}{\bibfnamefont{H.}~\bibnamefont{Saito}},
  \bibinfo{author}{\bibfnamefont{K.}~\bibnamefont{Uenishi}},
  \bibinfo{author}{\bibfnamefont{N.}~\bibnamefont{Miura}},
  \bibinfo{author}{\bibfnamefont{C.}~\bibnamefont{Tabata}},
  \bibinfo{author}{\bibfnamefont{H.}~\bibnamefont{Hidaka}},
  \bibinfo{author}{\bibfnamefont{T.}~\bibnamefont{Yanagisawa}},
  \bibnamefont{and} \bibinfo{author}{\bibfnamefont{H.}~\bibnamefont{Amitsuka}},
  \bibinfo{journal}{J. Phys. Soc. Jpn.} \textbf{\bibinfo{volume}{87}},
  \bibinfo{pages}{033702} (\bibinfo{year}{2018}).

\bibitem[{\citenamefont{Yanagisawa et~al.}(2021)\citenamefont{Yanagisawa,
  Matsumori, Saito, Hidaka, Amitsuka, Nakamura, Awaji, Gorbunov, Zherlitsyn,
  Wosnitza et~al.}}]{Yanagisawa_PhysRevLett.126.157201}
\bibinfo{author}{\bibfnamefont{T.}~\bibnamefont{Yanagisawa}},
  \bibinfo{author}{\bibfnamefont{H.}~\bibnamefont{Matsumori}},
  \bibinfo{author}{\bibfnamefont{H.}~\bibnamefont{Saito}},
  \bibinfo{author}{\bibfnamefont{H.}~\bibnamefont{Hidaka}},
  \bibinfo{author}{\bibfnamefont{H.}~\bibnamefont{Amitsuka}},
  \bibinfo{author}{\bibfnamefont{S.}~\bibnamefont{Nakamura}},
  \bibinfo{author}{\bibfnamefont{S.}~\bibnamefont{Awaji}},
  \bibinfo{author}{\bibfnamefont{D.~I.} \bibnamefont{Gorbunov}},
  \bibinfo{author}{\bibfnamefont{S.}~\bibnamefont{Zherlitsyn}},
  \bibinfo{author}{\bibfnamefont{J.}~\bibnamefont{Wosnitza}},
  \bibnamefont{et~al.}, \bibinfo{journal}{Phys. Rev. Lett.}
  \textbf{\bibinfo{volume}{126}}, \bibinfo{pages}{157201}
  (\bibinfo{year}{2021}).

\bibitem[{\citenamefont{Ota et~al.}(2022)\citenamefont{Ota, Shimozawa, Muroya,
  Miyamoto, Hosoi, Nakamura, Homma, Honda, Aoki, and Izawa}}]{ota2022zero}
\bibinfo{author}{\bibfnamefont{K.}~\bibnamefont{Ota}},
  \bibinfo{author}{\bibfnamefont{M.}~\bibnamefont{Shimozawa}},
  \bibinfo{author}{\bibfnamefont{T.}~\bibnamefont{Muroya}},
  \bibinfo{author}{\bibfnamefont{T.}~\bibnamefont{Miyamoto}},
  \bibinfo{author}{\bibfnamefont{S.}~\bibnamefont{Hosoi}},
  \bibinfo{author}{\bibfnamefont{A.}~\bibnamefont{Nakamura}},
  \bibinfo{author}{\bibfnamefont{Y.}~\bibnamefont{Homma}},
  \bibinfo{author}{\bibfnamefont{F.}~\bibnamefont{Honda}},
  \bibinfo{author}{\bibfnamefont{D.}~\bibnamefont{Aoki}}, \bibnamefont{and}
  \bibinfo{author}{\bibfnamefont{K.}~\bibnamefont{Izawa}},
  \bibinfo{journal}{arXiv:2205.05555}  (\bibinfo{year}{2022}).

\bibitem[{\citenamefont{Motoyama et~al.}(2018)\citenamefont{Motoyama, Sezaki,
  Gouchi, Miyoshi, Nishigori, Mutou, Fujiwara, and
  Uwatoko}}]{motoyama2018magnetic}
\bibinfo{author}{\bibfnamefont{G.}~\bibnamefont{Motoyama}},
  \bibinfo{author}{\bibfnamefont{M.}~\bibnamefont{Sezaki}},
  \bibinfo{author}{\bibfnamefont{J.}~\bibnamefont{Gouchi}},
  \bibinfo{author}{\bibfnamefont{K.}~\bibnamefont{Miyoshi}},
  \bibinfo{author}{\bibfnamefont{S.}~\bibnamefont{Nishigori}},
  \bibinfo{author}{\bibfnamefont{T.}~\bibnamefont{Mutou}},
  \bibinfo{author}{\bibfnamefont{K.}~\bibnamefont{Fujiwara}}, \bibnamefont{and}
  \bibinfo{author}{\bibfnamefont{Y.}~\bibnamefont{Uwatoko}},
  \bibinfo{journal}{Physica B: Condens. Matter} \textbf{\bibinfo{volume}{536}},
  \bibinfo{pages}{142} (\bibinfo{year}{2018}).

\bibitem[{\citenamefont{Shinozaki
  et~al.}(2020{\natexlab{a}})\citenamefont{Shinozaki, Motoyama, Tsubouchi,
  Sezaki, Gouchi, Nishigori, Mutou, Yamaguchi, Fujiwara, Miyoshi
  et~al.}}]{shinozaki2020magnetoelectric}
\bibinfo{author}{\bibfnamefont{M.}~\bibnamefont{Shinozaki}},
  \bibinfo{author}{\bibfnamefont{G.}~\bibnamefont{Motoyama}},
  \bibinfo{author}{\bibfnamefont{M.}~\bibnamefont{Tsubouchi}},
  \bibinfo{author}{\bibfnamefont{M.}~\bibnamefont{Sezaki}},
  \bibinfo{author}{\bibfnamefont{J.}~\bibnamefont{Gouchi}},
  \bibinfo{author}{\bibfnamefont{S.}~\bibnamefont{Nishigori}},
  \bibinfo{author}{\bibfnamefont{T.}~\bibnamefont{Mutou}},
  \bibinfo{author}{\bibfnamefont{A.}~\bibnamefont{Yamaguchi}},
  \bibinfo{author}{\bibfnamefont{K.}~\bibnamefont{Fujiwara}},
  \bibinfo{author}{\bibfnamefont{K.}~\bibnamefont{Miyoshi}},
  \bibnamefont{et~al.}, \bibinfo{journal}{J. Phys. Soc. Jpn.}
  \textbf{\bibinfo{volume}{89}}, \bibinfo{pages}{033703}
  (\bibinfo{year}{2020}{\natexlab{a}}).

\bibitem[{\citenamefont{Shinozaki
  et~al.}(2020{\natexlab{b}})\citenamefont{Shinozaki, Motoyama, Mutou,
  Nishigori, Yamaguchi, Fujiwara, Miyoshi, and Sumiyama}}]{shinozaki2020study}
\bibinfo{author}{\bibfnamefont{M.}~\bibnamefont{Shinozaki}},
  \bibinfo{author}{\bibfnamefont{G.}~\bibnamefont{Motoyama}},
  \bibinfo{author}{\bibfnamefont{T.}~\bibnamefont{Mutou}},
  \bibinfo{author}{\bibfnamefont{S.}~\bibnamefont{Nishigori}},
  \bibinfo{author}{\bibfnamefont{A.}~\bibnamefont{Yamaguchi}},
  \bibinfo{author}{\bibfnamefont{K.}~\bibnamefont{Fujiwara}},
  \bibinfo{author}{\bibfnamefont{K.}~\bibnamefont{Miyoshi}}, \bibnamefont{and}
  \bibinfo{author}{\bibfnamefont{A.}~\bibnamefont{Sumiyama}},
  \bibinfo{journal}{JPS Conf. Proc.} \textbf{\bibinfo{volume}{30}},
  \bibinfo{pages}{011189} (\bibinfo{year}{2020}{\natexlab{b}}).

\bibitem[{\citenamefont{Hayami and
  Kusunose}(2022)}]{Hayami_doi:10.7566/JPSJ.91.123701}
\bibinfo{author}{\bibfnamefont{S.}~\bibnamefont{Hayami}} \bibnamefont{and}
  \bibinfo{author}{\bibfnamefont{H.}~\bibnamefont{Kusunose}},
  \bibinfo{journal}{J. Phys. Soc. Jpn.} \textbf{\bibinfo{volume}{91}},
  \bibinfo{pages}{123701} (\bibinfo{year}{2022}).

\bibitem[{\citenamefont{Hayami and Kusunose}(2018)}]{hayami2018microscopic}
\bibinfo{author}{\bibfnamefont{S.}~\bibnamefont{Hayami}} \bibnamefont{and}
  \bibinfo{author}{\bibfnamefont{H.}~\bibnamefont{Kusunose}},
  \bibinfo{journal}{J. Phys. Soc. Jpn.} \textbf{\bibinfo{volume}{87}},
  \bibinfo{pages}{033709} (\bibinfo{year}{2018}).

\bibitem[{\citenamefont{Kusunose et~al.}(2020)\citenamefont{Kusunose, Oiwa, and
  Hayami}}]{kusunose2020complete}
\bibinfo{author}{\bibfnamefont{H.}~\bibnamefont{Kusunose}},
  \bibinfo{author}{\bibfnamefont{R.}~\bibnamefont{Oiwa}}, \bibnamefont{and}
  \bibinfo{author}{\bibfnamefont{S.}~\bibnamefont{Hayami}},
  \bibinfo{journal}{J. Phys. Soc. Jpn.} \textbf{\bibinfo{volume}{89}},
  \bibinfo{pages}{104704} (\bibinfo{year}{2020}).

\bibitem[{\citenamefont{Yatsushiro et~al.}(2021)\citenamefont{Yatsushiro,
  Kusunose, and Hayami}}]{Yatsushiro_PhysRevB.104.054412}
\bibinfo{author}{\bibfnamefont{M.}~\bibnamefont{Yatsushiro}},
  \bibinfo{author}{\bibfnamefont{H.}~\bibnamefont{Kusunose}}, \bibnamefont{and}
  \bibinfo{author}{\bibfnamefont{S.}~\bibnamefont{Hayami}},
  \bibinfo{journal}{Phys. Rev. B} \textbf{\bibinfo{volume}{104}},
  \bibinfo{pages}{054412} (\bibinfo{year}{2021}).

\bibitem[{\citenamefont{Kusunose et~al.}(2023)\citenamefont{Kusunose, Oiwa, and
  Hayami}}]{Kusunose_PhysRevB.107.195118}
\bibinfo{author}{\bibfnamefont{H.}~\bibnamefont{Kusunose}},
  \bibinfo{author}{\bibfnamefont{R.}~\bibnamefont{Oiwa}}, \bibnamefont{and}
  \bibinfo{author}{\bibfnamefont{S.}~\bibnamefont{Hayami}},
  \bibinfo{journal}{Phys. Rev. B} \textbf{\bibinfo{volume}{107}},
  \bibinfo{pages}{195118} (\bibinfo{year}{2023}).

\bibitem[{\citenamefont{Hayami et~al.}(2018{\natexlab{b}})\citenamefont{Hayami,
  Yatsushiro, Yanagi, and Kusunose}}]{Hayami_PhysRevB.98.165110}
\bibinfo{author}{\bibfnamefont{S.}~\bibnamefont{Hayami}},
  \bibinfo{author}{\bibfnamefont{M.}~\bibnamefont{Yatsushiro}},
  \bibinfo{author}{\bibfnamefont{Y.}~\bibnamefont{Yanagi}}, \bibnamefont{and}
  \bibinfo{author}{\bibfnamefont{H.}~\bibnamefont{Kusunose}},
  \bibinfo{journal}{Phys. Rev. B} \textbf{\bibinfo{volume}{98}},
  \bibinfo{pages}{165110} (\bibinfo{year}{2018}{\natexlab{b}}).

\bibitem[{\citenamefont{Oiwa and
  Kusunose}(2022)}]{Oiwa_doi:10.7566/JPSJ.91.014701}
\bibinfo{author}{\bibfnamefont{R.}~\bibnamefont{Oiwa}} \bibnamefont{and}
  \bibinfo{author}{\bibfnamefont{H.}~\bibnamefont{Kusunose}},
  \bibinfo{journal}{J. Phys. Soc. Jpn.} \textbf{\bibinfo{volume}{91}},
  \bibinfo{pages}{014701} (\bibinfo{year}{2022}).

\bibitem[{com()}]{comment_essential_sl}
\bibinfo{note}{It is noted that $c_i$ and $F$ are different for $\langle s^{\rm
  S} \rangle$ and $\langle s^{\rm AS} \rangle$ ($\langle l^{\rm S} \rangle$ and
  $\langle l^{\rm AS} \rangle$).}

\bibitem[{\citenamefont{Thiede et~al.}(1998)\citenamefont{Thiede, Ebel, and
  Jeitschko}}]{thiede1998ternary}
\bibinfo{author}{\bibfnamefont{V.~T.} \bibnamefont{Thiede}},
  \bibinfo{author}{\bibfnamefont{T.}~\bibnamefont{Ebel}}, \bibnamefont{and}
  \bibinfo{author}{\bibfnamefont{W.}~\bibnamefont{Jeitschko}},
  \bibinfo{journal}{J. Mat. Chem.} \textbf{\bibinfo{volume}{8}},
  \bibinfo{pages}{125} (\bibinfo{year}{1998}).

\bibitem[{\citenamefont{Reehuis et~al.}(2003)\citenamefont{Reehuis, Wolff,
  Krimmel, Scheidt, St{\"u}sser, Loidl, and Jeitschko}}]{reehuis2003magnetic}
\bibinfo{author}{\bibfnamefont{M.}~\bibnamefont{Reehuis}},
  \bibinfo{author}{\bibfnamefont{M.}~\bibnamefont{Wolff}},
  \bibinfo{author}{\bibfnamefont{A.}~\bibnamefont{Krimmel}},
  \bibinfo{author}{\bibfnamefont{E.}~\bibnamefont{Scheidt}},
  \bibinfo{author}{\bibfnamefont{N.}~\bibnamefont{St{\"u}sser}},
  \bibinfo{author}{\bibfnamefont{A.}~\bibnamefont{Loidl}}, \bibnamefont{and}
  \bibinfo{author}{\bibfnamefont{W.}~\bibnamefont{Jeitschko}},
  \bibinfo{journal}{J. Phys.: Condens. Matter} \textbf{\bibinfo{volume}{15}},
  \bibinfo{pages}{1773} (\bibinfo{year}{2003}).

\bibitem[{\citenamefont{Khalyavin et~al.}(2010)\citenamefont{Khalyavin,
  Hillier, Adroja, Strydom, Manuel, Chapon, Peratheepan, Knight, Deen, Ritter
  et~al.}}]{KhalyavinPhysRevB.82.100405}
\bibinfo{author}{\bibfnamefont{D.~D.} \bibnamefont{Khalyavin}},
  \bibinfo{author}{\bibfnamefont{A.~D.} \bibnamefont{Hillier}},
  \bibinfo{author}{\bibfnamefont{D.~T.} \bibnamefont{Adroja}},
  \bibinfo{author}{\bibfnamefont{A.~M.} \bibnamefont{Strydom}},
  \bibinfo{author}{\bibfnamefont{P.}~\bibnamefont{Manuel}},
  \bibinfo{author}{\bibfnamefont{L.~C.} \bibnamefont{Chapon}},
  \bibinfo{author}{\bibfnamefont{P.}~\bibnamefont{Peratheepan}},
  \bibinfo{author}{\bibfnamefont{K.}~\bibnamefont{Knight}},
  \bibinfo{author}{\bibfnamefont{P.}~\bibnamefont{Deen}},
  \bibinfo{author}{\bibfnamefont{C.}~\bibnamefont{Ritter}},
  \bibnamefont{et~al.}, \bibinfo{journal}{Phys. Rev. B}
  \textbf{\bibinfo{volume}{82}}, \bibinfo{pages}{100405}
  (\bibinfo{year}{2010}).

\bibitem[{\citenamefont{Tanida et~al.}(2011)\citenamefont{Tanida, Tanaka, Sera,
  Tanimoto, Nishioka, Matsumura, Ogawa, Moriyoshi, Kuroiwa, Kim
  et~al.}}]{TanidaPhysRevB.84.115128}
\bibinfo{author}{\bibfnamefont{H.}~\bibnamefont{Tanida}},
  \bibinfo{author}{\bibfnamefont{D.}~\bibnamefont{Tanaka}},
  \bibinfo{author}{\bibfnamefont{M.}~\bibnamefont{Sera}},
  \bibinfo{author}{\bibfnamefont{S.}~\bibnamefont{Tanimoto}},
  \bibinfo{author}{\bibfnamefont{T.}~\bibnamefont{Nishioka}},
  \bibinfo{author}{\bibfnamefont{M.}~\bibnamefont{Matsumura}},
  \bibinfo{author}{\bibfnamefont{M.}~\bibnamefont{Ogawa}},
  \bibinfo{author}{\bibfnamefont{C.}~\bibnamefont{Moriyoshi}},
  \bibinfo{author}{\bibfnamefont{Y.}~\bibnamefont{Kuroiwa}},
  \bibinfo{author}{\bibfnamefont{J.~E.} \bibnamefont{Kim}},
  \bibnamefont{et~al.}, \bibinfo{journal}{Phys. Rev. B}
  \textbf{\bibinfo{volume}{84}}, \bibinfo{pages}{115128}
  (\bibinfo{year}{2011}).

\bibitem[{\citenamefont{Mignot et~al.}(2011)\citenamefont{Mignot, Robert,
  Andr{\'e}, M.~Bataille, Nishioka, Kobayashi, Matsumura, Tanida, Tanaka, and
  Sera}}]{mignot2011neutron}
\bibinfo{author}{\bibfnamefont{J.-M.} \bibnamefont{Mignot}},
  \bibinfo{author}{\bibfnamefont{J.}~\bibnamefont{Robert}},
  \bibinfo{author}{\bibfnamefont{G.}~\bibnamefont{Andr{\'e}}},
  \bibinfo{author}{\bibfnamefont{A.}~\bibnamefont{M.~Bataille}},
  \bibinfo{author}{\bibfnamefont{T.}~\bibnamefont{Nishioka}},
  \bibinfo{author}{\bibfnamefont{R.}~\bibnamefont{Kobayashi}},
  \bibinfo{author}{\bibfnamefont{M.}~\bibnamefont{Matsumura}},
  \bibinfo{author}{\bibfnamefont{H.}~\bibnamefont{Tanida}},
  \bibinfo{author}{\bibfnamefont{D.}~\bibnamefont{Tanaka}}, \bibnamefont{and}
  \bibinfo{author}{\bibfnamefont{M.}~\bibnamefont{Sera}}, \bibinfo{journal}{J.
  Phys. Soc. Jpn.} \textbf{\bibinfo{volume}{80}}, \bibinfo{pages}{SA022}
  (\bibinfo{year}{2011}).

\bibitem[{\citenamefont{Muro et~al.}(2011)\citenamefont{Muro, Kajino, Onimaru,
  and Takabatake}}]{muro2011magnetic}
\bibinfo{author}{\bibfnamefont{Y.}~\bibnamefont{Muro}},
  \bibinfo{author}{\bibfnamefont{J.}~\bibnamefont{Kajino}},
  \bibinfo{author}{\bibfnamefont{T.}~\bibnamefont{Onimaru}}, \bibnamefont{and}
  \bibinfo{author}{\bibfnamefont{T.}~\bibnamefont{Takabatake}},
  \bibinfo{journal}{J. Phys. Soc. Jpn.} \textbf{\bibinfo{volume}{80}},
  \bibinfo{pages}{SA021} (\bibinfo{year}{2011}).

\bibitem[{\citenamefont{Kato et~al.}(2011)\citenamefont{Kato, Kobayashi,
  Takesaka, Nishioka, Matsumura, Kaneko, and Metoki}}]{kato2011magnetic}
\bibinfo{author}{\bibfnamefont{H.}~\bibnamefont{Kato}},
  \bibinfo{author}{\bibfnamefont{R.}~\bibnamefont{Kobayashi}},
  \bibinfo{author}{\bibfnamefont{T.}~\bibnamefont{Takesaka}},
  \bibinfo{author}{\bibfnamefont{T.}~\bibnamefont{Nishioka}},
  \bibinfo{author}{\bibfnamefont{M.}~\bibnamefont{Matsumura}},
  \bibinfo{author}{\bibfnamefont{K.}~\bibnamefont{Kaneko}}, \bibnamefont{and}
  \bibinfo{author}{\bibfnamefont{N.}~\bibnamefont{Metoki}},
  \bibinfo{journal}{J. Phys. Soc. Jpn.} \textbf{\bibinfo{volume}{80}}
  (\bibinfo{year}{2011}).

\bibitem[{\citenamefont{Li et~al.}(2019)\citenamefont{Li, Honda, Miyake, Homma,
  Haga, Nakamura, Shimizu, Maurya, Sato, Tokunaga
  et~al.}}]{Li_PhysRevB.99.054408}
\bibinfo{author}{\bibfnamefont{D.~X.} \bibnamefont{Li}},
  \bibinfo{author}{\bibfnamefont{F.}~\bibnamefont{Honda}},
  \bibinfo{author}{\bibfnamefont{A.}~\bibnamefont{Miyake}},
  \bibinfo{author}{\bibfnamefont{Y.}~\bibnamefont{Homma}},
  \bibinfo{author}{\bibfnamefont{Y.}~\bibnamefont{Haga}},
  \bibinfo{author}{\bibfnamefont{A.}~\bibnamefont{Nakamura}},
  \bibinfo{author}{\bibfnamefont{Y.}~\bibnamefont{Shimizu}},
  \bibinfo{author}{\bibfnamefont{A.}~\bibnamefont{Maurya}},
  \bibinfo{author}{\bibfnamefont{Y.~J.} \bibnamefont{Sato}},
  \bibinfo{author}{\bibfnamefont{M.}~\bibnamefont{Tokunaga}},
  \bibnamefont{et~al.}, \bibinfo{journal}{Phys. Rev. B}
  \textbf{\bibinfo{volume}{99}}, \bibinfo{pages}{054408}
  (\bibinfo{year}{2019}).

\end{thebibliography}
\end{document}